\documentclass[11pt]{article}

\usepackage{indentfirst}
\usepackage{amsmath}
\usepackage{amsthm}
\usepackage{color}
\usepackage{eufrak}
\usepackage{verbatim}
\usepackage[makeroom]{cancel}
\usepackage{rotating}
\usepackage{float}
\usepackage[margin=1.15in]{geometry}
\usepackage{amssymb}
\usepackage{amsthm}
\usepackage{mathrsfs}
\usepackage{mathtools}
\usepackage{epsfig}
\usepackage{graphicx}
\usepackage{indentfirst}
\usepackage{framed}
\usepackage[numbers]{natbib}
\usepackage{color}
\usepackage{hyperref}
\usepackage{graphicx}
\usepackage{hyperref}
\usepackage{float}
\usepackage{bm}
\usepackage{cleveref}
\usepackage[labelfont=bf]{caption}
\usepackage{algorithm}
\usepackage[noend]{algpseudocode}
\usepackage{soul}

\definecolor{corlinks}{RGB}{0,0,150}
\definecolor{cormenu}{RGB}{0,0,150}
\definecolor{corurl}{RGB}{0,0,150}

\hypersetup{
	colorlinks=true,
	urlcolor=corlinks,
	linkcolor=corlinks,
	menucolor=cormenu,
	citecolor=corlinks,
	pdfborder= 0 0 0
}

\newtheorem{theorem}{Theorem}

\newtheorem{lemma}[theorem]{Lemma}
\newtheorem{corollary}[theorem]{Corollary}
\newtheorem{definition}[theorem]{Definition}
\newtheorem{proposition}[theorem]{Proposition}
\newtheorem{remark}[theorem]{Remark}

\newtheorem{fact}[theorem]{Fact}
\newtheorem{claim}[theorem]{Claim}

\newtheorem{hypothesis}[theorem]{Hypothesis}

\DeclareMathOperator{\poly}{poly}

\DeclareMathOperator*\Exp{{\bf E}}
\DeclareMathOperator*\Prob{{\bf Pr}}
\newcommand{\bool}{\left\{0,1\right\}}
\newcommand{\Kt}{\mathsf{Kt}}
\newcommand{\rKt}{\mathsf{rKt}}

\newcommand{\mcX}{\mathcal X}
\newcommand{\mcY}{\mathcal Y}

\newcommand{\comp}{\mathsf{Compress}}
\newcommand{\dcomp}{\mathsf{Decompress}}
\newcommand{\CryptoETH}{$\gamma$-$\mathsf{Crypto}$-$\mathsf{ETH}$}

\newcommand{\K}{\mathsf{K}}
\newcommand{\rK}{\mathsf{rK}}
\newcommand{\pK}{\mathsf{pK}}

\newcommand{\eqdef}{\stackrel{\rm def}{=}}

\newcommand{\BPP}{\mathsf{BPP}}

\newcommand{\NP}{\mathsf{NP}}
\newcommand{\PH}{\mathsf{PH}}

\newcommand{\rc}{\rho_C}
\newcommand{\rd}{\rho_D}
\newcommand{\eps}{\varepsilon}

\newcommand{\tm}{t_{max}}
\newcommand{\guv}{\textrm{GUV}}
\newcommand{\bz}{\textrm{BZ}}

\def\caD{\mathcal{D}}
\def\caH{\mathcal{H}}

\def\caM{\mathcal{M}}

\usepackage{algorithm}
\usepackage[noend]{algpseudocode}

\definecolor{blue-violet}{rgb}{0.54, 0.17, 0.89}
\definecolor{darkorange}{rgb}{1.0, 0.55, 0.0}

\newcommand{\EXP}{\mathsf{EXP}}
\renewcommand{\P}{\mathsf{P}}

\newcommand{\zo}{\{0,1\}}
\newcommand{\mapping}{\rightarrow}
\newcommand{\ff}{\mathrm{F}}

\newcommand{\bnote}[1]{\textcolor{red}{(Bruno: #1)}}

\begin{document}

	\newgeometry{margin=0.91in}

	\title{Optimal Coding Theorems in Time-Bounded Kolmogorov Complexity\vspace{0.4cm}}

	
	\author{Zhenjian Lu\thanks{University of Warwick, UK. \texttt{E-mail:~zhen.j.lu@warwick.ac.uk}}\and
	Igor C. Oliveira\thanks{University of Warwick, UK. \texttt{E-mail:~igor.oliveira@warwick.ac.uk}} \and Marius Zimand\thanks{Towson University, US. \texttt{E-mail:~mzimand@towson.edu}.}\vspace{0.5cm}}
	
	\date{}  
	
	\maketitle
	
	\vspace{-0.45cm}

	\begin{abstract}
			The classical coding theorem in Kolmogorov complexity states that if an $n$-bit string $x$ is sampled with probability $\delta$ by an algorithm with prefix-free domain then $\mathsf{K}(x) \leq \log(1/\delta) + O(1)$. In a recent work, Lu and Oliveira \cite{DBLP:conf/icalp/LuO21} established an unconditional time-bounded version of this result, by showing that if $x$ can be efficiently sampled with probability $\delta$ then $\rKt(x) = O(\log(1/\delta)) + O(\log n)$, where $\rKt$ denotes the randomized analogue of Levin's $\Kt$ complexity. Unfortunately, this result is often insufficient when transferring applications of the classical coding theorem to the time-bounded setting, as it  achieves a  $O(\log(1/\delta))$ bound instead of the information-theoretic optimal $\log(1/\delta)$.
			
			Motivated by this discrepancy, we investigate optimal coding theorems in the time-bounded setting. Our main contributions can be summarised as follows.\\
			
			
		\noindent	\textbf{$\bullet$~Efficient coding theorem for $\mathsf{rKt}$ with a factor of $2$.} Addressing a question from \cite{DBLP:conf/icalp/LuO21}, we show that if $x$ can be efficiently sampled with probability at least $\delta$ then $\rKt(x) \le (2 + o(1)) \cdot \log(1/\delta) + O\!\left(\log n\right)$. As in previous work, our coding theorem is \emph{efficient} in the sense that it provides a polynomial-time probabilistic algorithm that, when given $x$, the code of the sampler, and $\delta$,  it outputs, with  probability $\ge 0.99$, a probabilistic representation of $x$ that certifies this $\rKt$ complexity bound.\\ 
			
		\noindent	    \textbf{$\bullet$~Optimality under a cryptographic assumption.} Under a hypothesis about the security of cryptographic pseudorandom generators, we show that no efficient coding theorem can achieve a bound of the form $\rKt(x) \leq (2 - o(1)) \cdot \log(1/\delta) + \mathsf{poly}(\log n)$. Under a weaker assumption, we exhibit a gap between \emph{efficient} coding theorems and  \emph{existential} coding theorems with near-optimal parameters.\\
		
			    
		\noindent	    \textbf{$\bullet$~Optimal coding theorem for $\pK^t$ and unconditional Antunes-Fortnow.} We consider $\pK^t$ complexity \cite{GKLO22}, a variant of $\rKt$ where the randomness is public and the time bound is fixed. We observe the existence of an optimal coding theorem for $\pK^t$, and employ this result to establish an  \emph{unconditional} version of a theorem of Antunes and Fortnow \cite{DBLP:conf/coco/AntunesF09} which characterizes the worst-case running times of languages that are in average polynomial-time over all $\mathsf{P}$-samplable distributions.

	\end{abstract}

		\restoregeometry

	\newpage

	\setcounter{tocdepth}{3}
	\tableofcontents
	
	\newpage

	\section{Introduction}

\subsection{Context and Background}

	A sampler is a probabilistic function that outputs Boolean strings. For any string $x \in \{0,1\}^*$ in its range, let $\mu(x)$ denote the probability with which $x$ is generated. The Coding Theorem in Kolmogorov complexity states that if the sampler is computable and its domain is a prefix-free set, then for every $x$ in its range  \[
	\K(x) \leq \log(1/\mu(x)) + O(1),
	\]
	where $\K(\cdot)$ is the prefix-free Kolmogorov complexity. In other words, strings that are sampled with non-trivial probability have short representations. Note that the coding theorem achieves \emph{optimal} expected length, since no uniquely decodable code can have expected length smaller than $\sum \mu(x) \log_2 (1/\mu(x))$, the entropy of the sampler (the sum is over all $x$ in the range of the sampler, assumed here to be finite). 
	
	The coding theorem is a central result in Kolmogorov complexity.\footnote{For instance, \cite{TroyLeeThesis} describes it as one of the four pillars of Kolmogorov complexity.} While it has found a number of applications in theoretical computer science (see, e.g., \cite{DBLP:journals/ipl/LiV92, TroyLeeThesis, DBLP:journals/mst/Aaronson14}), it comes with an important caveat: many aspects of the theory of Kolmogorov complexity are \emph{non-constructive}. For instance, there is provably no algorithm that estimates $\K(x)$. Similarly, for arbitrary samplers, there is no effective compressor achieving the short representation provided by the coding theorem\footnote{However, there exists a probabilistic polynomial-time compressor that given $x$ and an integer $m \geq \log (1/\mu(x))$  outputs a  description of $x$ of length $m$ + small polylogarithmic overhead~\cite{DBLP:journals/corr/abs-1911-04268}.} and also no upper bound on the running time required to decompress $x$ from it.
	
	In order to translate results and techniques from Kolmogorov complexity to the setting of \emph{efficient} algorithms and computations, several \emph{time-bounded} variants of Kolmogorov complexity have been proposed. We refer to the book~\cite{li-vit:b:kolmbook-four}, thesis~\cite{TroyLeeThesis}, and the surveys \cite{allender1992applications,  allender2001worlds, fortnow2004kolmogorov,   allender2017complexity} for a comprehensive treatment of this area and its numerous applications to algorithms, complexity, cryptography, learning, and pseudorandomness, among other fields. We highlight that 
	many exciting new results, which include worst-case to average-case reductions for NP problems \cite{DBLP:conf/focs/Hirahara18, DBLP:conf/stoc/Hirahara21} and complexity-theoretic characterizations of one-way functions \cite{DBLP:conf/focs/LiuP20, DBLP:conf/coco/RenS21}, rely in a crucial way on time-bounded Kolmogorov complexity. These recent developments further motivate the investigation of key results from Kolmogorov complexity in the time-bounded setting.

	In time-bounded Kolmogorov complexity we consider the minimum description length of a string $x$ \emph{with respect to machines that operate under a time constraint}.  We informally review next two central notions in this area (see Section \ref{sec:preliminaries} for precise definitions).  For a Turing machine $\caM$, we let $|\caM|$ denote its description length according to a fixed universal machine $U$. $\caM(\varepsilon)$ denotes the computation of $\caM$ over the empty string.\\
	 \vspace{-0.2cm}

\noindent \textbf{$\bm{\Kt}$ Complexity} \cite{DBLP:journals/iandc/Levin84}. This notion simultaneously considers description length and running time when measuring the complexity of a string $x$.
	\[
	\Kt(x) \;= \; \min_{\text{TM}\,\caM, \;t \geq 1} \left\{|\caM| + \log t \mid \caM(\varepsilon)~\text{outputs}~x~\text{in}~t~\text{steps}\right\}.
	\]

\noindent \textbf{$\bm{\K^t}$ Complexity} \cite{DBLP:conf/stoc/Sipser83a}. In contrast with $\Kt$, here we fix the time bound $t \colon \mathbb{N} \to \mathbb{N}$, and consider the minimum description with respect to machines that run in time at most $t(|x|)$.
	\[
	\K^t(x) \;= \; \min_{\text{TM}\,\caM} \left\{|\caM|  \,\mid\, \caM(\varepsilon)~\text{outputs}~x~\text{in}~t(|x|)~\text{steps}\right\}.
	\] 
While $\Kt$ complexity is tightly related to optimal search algorithms (see \cite{krajicek_2021} for a recent application), $\K^t$ is particularly useful in settings where maintaining a polynomial bound on the running time $t$ is desired (see, e.g., \cite{DBLP:conf/focs/Hirahara18}).

Antunes and Fortnow \cite{DBLP:conf/coco/AntunesF09} introduced techniques that can be used to establish  (conditional) coding theorems for $\K^t$ and $\Kt$. In particular, if a sampler runs in \emph{polynomial time} and outputs a string $x$ with probability at least $\delta$, then $\Kt(x) \leq \log(1/\delta) + O(\log n)$. Note that this coding theorem also achieves an optimal dependence on the probability parameter $\delta$. However, the results of \cite{DBLP:conf/coco/AntunesF09} rely on a strong derandomization assumption. For this reason, their application often lead to \emph{conditional} results.

More recently, \cite{DBLP:conf/icalp/LuO21} established an \emph{unconditional} coding theorem for a \emph{randomized} analogue of $\Kt$ complexity. Before explaining their result, we review the definitions of $\rKt$ and $\rK^t$.\\
 \vspace{-0.2cm}

\noindent \textbf{$\bm{\rKt}$ Complexity} \cite{DBLP:conf/icalp/Oliveira19}. In this definition, we consider randomized machines that output $x$ with high probability. 
    \[
	\rKt(x) \;= \; \min_{\text{RTM}\,\caM, \;t \geq 1} \left\{|\caM| + \log t \mid \caM(\varepsilon)~\text{outputs}~x~\text{in}~t~\text{steps with probability} \geq 2/3\right\}.
	\]

\noindent \textbf{$\bm{\rK^t}$ Complexity} \cite{DBLP:journals/cc/BuhrmanLM05, LOS21}.\footnote{\cite{DBLP:journals/cc/BuhrmanLM05} refers to this notion as $\mathsf{CBP}^t$ complexity.} This is the randomized analogue of $\K^t$, where the time bound $t$ is fixed in advance.
    \[
	\rK^t(x) \;= \; \min_{\text{RTM}\,\caM} \left\{|\caM|  \,\mid\, \caM(\varepsilon)~\text{outputs}~x~\text{in}~t(|x|)~\text{steps with probability} \geq 2/3\right\}.
	\]

In both cases, we can think of the randomized Turing machine $\caM$ as a \emph{probabilistic representation} of the input string $x$, in the sense that $x$ can be recovered with high probability from its description. These measures allow us to employ methods from time-bounded Kolmogorov complexity in the setting of randomized computation, which is ubiquitous in modern computer science. For instance, \cite{DBLP:conf/icalp/Oliveira19, LOS21} employed $\rKt$ and $\rK^t$ to obtain bounds on the compressibility of prime numbers and other objects and to show that certain problems about time-bounded Kolmogorov complexity can be intractable. We note that, under derandomization assumptions (see \cite{DBLP:conf/icalp/Oliveira19}), for every string $x$,  $\rKt(x) = \Theta(\Kt(x))$. Similarly, one can conditionally show that $\K^t(x)$ is essentially $\rK^t(x)$, up to a $O(\log |x|)$ additive term (see \cite{GKLO22}). Consequently, insights obtained in the context of probabilistic notions of Kolmogorov complexity can often inform the study of more classical notions such as $\Kt$ and $\K^t$.

Among other results, \cite{DBLP:conf/icalp/LuO21} established the following unconditional coding theorem in time-bounded Kolmogorov complexity: if a sampler runs in polynomial time and outputs a string $x$ with probability at least $\delta$, then $\rKt(x) = O(\log(1/\delta) + O(\log n)$. While this result can be used to port some applications of the coding theorem from Kolmogorov complexity to the time-bounded setting, in many cases it is still insufficient. This is because its dependence on the probability parameter $\delta$ is not optimal, which is often crucial in applications (see, e.g., \cite{DBLP:conf/coco/AntunesF09, DBLP:journals/mst/Aaronson14}).

	\subsection{Results}
	
	In this work, we investigate optimal coding theorems in time-bounded Kolmogorov complexity. We describe our results next.

\subsubsection{A Tighter Efficient Coding Theorem}\label{sec:intro_rkt_coding}
	
Our first result addresses the question posed in \cite[Problem 37]{DBLP:conf/icalp/LuO21}.

\begin{theorem}\label{thm:rkt_coding}
Suppose there is an efficient algorithm $A$ for sampling strings such that $A(1^n)$ outputs a string $x \in \{0,1\}^n$ with probability at least $\delta$. Then
\[
\rKt(x) \;\leq\; 2 \log(1/\delta) + O\!\left(\log n+ \log^2 \log(1/\delta)\right),
\]
where the constant behind the $O(\cdot)$ depends on $A$ and is independent of the remaining parameters. Moreover, given $x$, the code of $A$, and $\delta$, it is possible to compute in time $\mathsf{poly}(n,|A|)$, with probability $\geq 0.99$, a probabilistic representation of $x$ certifying this $\rKt$-complexity bound.
\end{theorem}

In \cite[Lemma 4]{DBLP:journals/siamcomp/BuhrmanFL01}, it was observed that by hashing modulo prime numbers one can obtain short descriptions of strings. As discussed in \cite[Section A.2.1]{DBLP:conf/icalp/LuO21}, for each efficient sampling algorithm, this technique implies that if some string $x$ is produced with probability $\geq \delta$, then  
$\rKt(x) \le 3\log (1/\delta) + O(\log n)$.\footnote{The bound from \cite[Section A.2.1]{DBLP:conf/icalp/LuO21} is different because it does not take into account the running time, which incurs an additional overhead of $\log(1/\delta)$.} In contrast, \Cref{thm:rkt_coding} achieves a bound of the form $(2 + o(1)) \cdot \log(1/\delta) + O(\log n)$. 

\Cref{thm:rkt_coding} readily improves some parameters in the applications of the coding theorem for $\rKt$ discussed in \cite{DBLP:conf/icalp/LuO21}, such as the efficient instance-based search-to-decision reduction for $\rKt$. We omit the details. 

In Section \ref{sec:efficient_rkt_coding}, we discuss extensions of this result. In particular, we describe precise bounds on the running time used in producing the corresponding probabilistic representation, and discuss computational aspects of the compression and decompression of $x$ in detail. In \Cref{s:estimate}, we discuss the computation of a probabilistic representation of the string $x$ when one does not know a probability bound $\delta$.

\subsubsection{Matching Lower Bound Under a Cryptographic Assumption}\label{sec:intro_lb}

It is possible to extend techniques from \cite{DBLP:conf/coco/AntunesF09} to show the following conditional result (see Section \ref{sec:existential_coding}).

\begin{proposition}\label{p:conditional_optimal_coding_rkt}
 Assume there is a language $L\in {\sf BPTIME}\left[2^{O(n)}\right]$ that requires nondeterministic circuits of size $2^{\Omega(n)}$ for all but finitely many $n$. Suppose there is an efficient algorithm $A$ for sampling strings such that $A(1^n)$ outputs a string $x \in \{0,1\}^n$ with probability at least $\delta > 0$. Then
\[
\rKt(x) \;\leq\; \log (1/\delta) + O(\log n). 
\]
\end{proposition}

While \Cref{p:conditional_optimal_coding_rkt} provides a better bound than \Cref{thm:rkt_coding}, the result is only \emph{existential}, i.e., it does not provide an efficient algorithm that produces a probabilistic representation of $x$. In other words, \Cref{p:conditional_optimal_coding_rkt} does not establish an \emph{efficient} coding theorem. Our next result shows that the bound achieved by \Cref{thm:rkt_coding} is optimal for efficient coding theorems, under a cryptographic assumption.\\

\vspace{-0.2cm}

\noindent \textbf{The Cryptographic Assumption.} For a constant $\gamma \in (0,1)$, we introduce the \CryptoETH~assumption, which can be seen as a cryptographic analogue of the well-known exponential time hypothesis about the complexity of $k$-CNF SAT \cite{DBLP:journals/jcss/ImpagliazzoP01}. Informally, we say that \CryptoETH~holds if there is a pseudorandom generator $G \colon \{0,1\}^{\ell(n)} \to \{0,1\}^n$ computable in time $\poly(n)$ that fools {\em uniform algorithms} running in time $2^{\gamma \cdot \ell(n)}$. Any seed length $(\log n)^{\omega(1)} \leq \ell(n) \leq n/2$ is sufficient in our negative results.  

In analogy with the well-known $\mathsf{ETH}$ and $\mathsf{SETH}$ hypotheses about the complexity of $k$-CNF SAT, we say that  $\mathsf{Crypto}\text{-}\mathsf{ETH}$ holds if \CryptoETH~is true for some $\gamma > 0$, and that $\mathsf{Crypto}\text{-}\mathsf{SETH}$ holds if \CryptoETH~is true for every $\gamma  \in (0,1)$. Since a candidate PRG of seed length $\ell(n)$ can be broken in time $2^{\ell(n)} \mathsf{poly}(n)$ by trying all possible seeds, these hypotheses postulate that for some PRGs one cannot have an attack that does sufficiently better than this naive brute-force approach. 

We stress that these assumptions refer to uniform algorithms. In the case of non-uniform distinguishers, it is known that $\mathsf{Crypto}\text{-}\mathsf{SETH}$ does not hold (see \cite{DBLP:journals/siamcomp/FiatN99, DBLP:conf/crypto/DeTT10, DBLP:conf/focs/ChungGLQ20} and references therein). We provide a formal treatment of the  cryptographic assumption in~\Cref{sec:cond_lower_bound}.

\begin{theorem}[Informal]\label{thm:coding_lower_bound}
Let $\gamma \in (0,1)$ be any constant. If \CryptoETH~holds, there is no efficient coding theorem for $\rKt$ that achieves bounds of the form $(1 + \gamma - o(1)) \cdot \log(1/\delta) + \mathsf{poly}(\log n)$.
\end{theorem}

	\Cref{thm:coding_lower_bound} shows that if $\mathsf{Crypto}\text{-}\mathsf{ETH}$ holds then the best parameter achieved by an \emph{efficient} coding theorem for $\rKt$ is $(1 + \Omega(1)) \cdot \log(1/\delta) + \mathsf{poly}(\log n)$. This exhibits an inherent gap in parameters between the efficient coding theorem (\Cref{thm:rkt_coding}) and its existential analogue (\Cref{p:conditional_optimal_coding_rkt}). On the other hand, if the stronger $\mathsf{Crypto}\text{-}\mathsf{SETH}$ hypothesis holds, then no efficient coding theorem for $\rKt$ achieves parameter $(2 - o(1)) \cdot \log(1/\delta) + \mathsf{poly}(\log n)$. In this case, \Cref{thm:rkt_coding} is essentially optimal with respect to its dependence on $\delta$.\\
	
	\vspace{-0.2cm}

\noindent \textbf{Fine-grained complexity of coding algorithms for polynomial-time samplers.} An $\rKt$ bound refers to the time necessary to \emph{decompress} a string $x$ from its probabilistic representation. On the other hand, an \emph{efficient} coding theorem provides a routine that can \emph{compress} $x$ in polynomial time. More generally, a coding procedure for a sampler $A$ consists of a pair of probabilistic algorithms $(\comp, \dcomp)$ that aim to produce a ``good'' codeword $p$ for every string $y$ sampled by $A$.  The quality of $p$ depends on three values: the length of $p$, the number of steps $t_C$ used to produce $p$ from $y$ (the compression time), and the number of steps $t_D$ used to produce $y$ from $p$ (the decompression time). It is interesting to understand the trade-off between these three values. Toward this goal, we aggregate them in a manner similar to $\rKt$, by defining 
the 2-sided-$\rKt$ complexity of $y$ to be, roughly, $|p| + \log (t_C + t_D)$ (the formal~\Cref{d:boundcompress} is more complicated because it takes into account that $\comp$ and $\dcomp$ are probabilistic). Thus according to 2-sided-$\rKt$, each bit gained by a shorter codeword is worth doubling the compression/decompression time. For instance, for simple samplers (say, having a finite range, or generating strings with the uniform distribution),  there exist trivial polynomial time $\comp$ and $\dcomp$, which in case $y$ is sampled with probability at least $\delta$, produce  a codeword $p$ with $|p| = \log(1/\delta)$ (provided $\comp$ and $\dcomp$ know $\delta$). Such a coding procedure certifies for each sampled string  a 2-sided-$\rKt$ complexity of $\log(1/\delta) + O(\log n)$. We say that the sampler admits coding with $2$-sided-$\rKt$ complexity bounded by  $\log(1/\delta) + O(\log n)$. In general, we have to include also the error probability  of  $\comp$ and $\dcomp$, which we omit in this informal discussion.
 
Similarly to~\Cref{thm:rkt_coding} and~\Cref{thm:coding_lower_bound} (and also with similar proofs), we establish the following theorem. 
\begin{theorem}[Informal]\label{t:twosidedinformal} The following results hold.
\begin{itemize}
    \item[\emph{(}a\emph{)}] \emph{(Upper Bound)} Every polynomial-time sampler admits  coding with $2$-sided-$\rKt$ complexity $2 \log (1/\delta) + O(\log^2 \log (1/\delta)) + O(\log n)$.
    \item[\emph{(}b\emph{)}] \emph{(Conditional Lower Bound)} Let $\gamma \in (0,1)$ be any constant. If \CryptoETH~holds, there  exists a polynomial-time sampler that  does not admit coding with $2$-sided-$\rKt$ complexity bounded by  $(1 + \gamma - o(1)) \cdot \log(1/\delta) + \mathsf{poly}(\log n)$, unless the error probability is greater than $1/7$.
\end{itemize}
\end{theorem}

\subsubsection{An Optimal Coding Theorem and Unconditional Antunes-Fortnow}

While \Cref{thm:rkt_coding} improves the result from \cite{DBLP:conf/icalp/LuO21} to achieve a bound that is tight up to a factor of $2$ and that is possibly optimal among efficient coding theorems, it is still insufficient in many applications. We consider next a variant of $\rKt$ that allows us to establish an \emph{optimal} and \emph{unconditional} coding theorem in time-bounded Kolmogorov complexity.

Fix a function $t \colon \mathbb{N} \to \mathbb{N}$. For a string $x\in\bool^*$, the probabilistic $t$-bounded Kolmogorov complexity of $x$ (see \cite{GKLO22}) is defined as
		\[
		\pK^{t}(x) = \min\left\{k \,\,\,\middle\vert\,\,\, \Prob_{w\sim\bool^{t(|x|)}}\left[\text{$\exists\, \caM\in\bool^{k}$,  $\caM(w)$  outputs $x$ within $t(|x|)$ steps} \right] \geq \frac{2}{3}\right\}.
		\]
In other words, if $k = \pK^{t}(x)$, then with probability  at least $2/3$ over the choice of the random string $w$, $x$ admits a time-bounded encoding of length $k$. In particular, if two parties share a typical random string $w$, then $x$ can be transmitted with $k$ bits and decompressed in time $t = t(|x|)$. (Recall that here the time bound $t$ is fixed, as opposed to $\rKt$, where a $\log t$ term is added to the description length.) 

It is possible to show that $\K^{t}(x)$, $\rK^t(x)$, and $\pK^t(x)$ correspond essentially to the same time-bounded measure, under standard derandomization assumptions \cite{GKLO22}.\footnote{More precisely, under standard derandomization assumptions, $\pK^{t}(x)$ and $\rK^{t'}(x)$ coincide up to an additive term of $O(\log |x|)$, provided that $t' = \poly(t)$. A similar relation holds between $\K^t$ and $\rK^t$.} One of the main benefits of $\pK^t$ is that it allows us to establish unconditional results that are currently unknown in the case of the other measures.\footnote{While in this work we focus on coding theorems, we stress that $\pK^t$ is a key notion introduced in \cite{GKLO22} that enables the investigation of meta-complexity in the setting of probabilistic computations. It has applications in worst-case to average-case reductions and in learning theory.}

\begin{theorem}\label{thm:pkt_coding}
Suppose there is a randomized algorithm $A$ for sampling strings such that $A(1^n)$ runs in time $T(n) \geq n$ and outputs a string $x \in \{0,1\}^n$ with probability at least $\delta > 0$. Then
		\[
		\pK^t(x) \,=\,  \log(1/\delta) +  O\!\left(\log T(n)\right),
		\]
		where $t(n) =\poly\!\left(T(n)\right)$ and the constant behind the $O(\cdot)$ depends on $|A|$ and is independent of the remaining parameters.
\end{theorem}

\Cref{thm:pkt_coding} provides a time-bounded coding theorem that can be used in settings where the optimal dependence on $\delta$ is crucial. As an immediate application, it is possible to show an equivalence between efficiently sampling a fixed sequence $w_n \in \{0,1\}^n$ of objects (e.g.,~$n$-bit prime numbers) with probability at least $\delta_n/\mathsf{poly}(n)$ and the existence of bounds for the corresponding objects of the form $\pK^{\poly}(w_n) = \log(1/\delta_n) + O(\log n)$.\footnote{An efficient sampler immediately implies the corresponding $\pK^t$ bounds via \Cref{thm:pkt_coding}. On the other hand, objects of bounded $\pK^t$ complexity can be sampled by considering a random sequence of bits and a random program of appropriate length. We refer to \cite[Theorem 6]{DBLP:conf/icalp/LuO21} for a weaker relation and its proof. Since the argument is essentially the same, we omit the precise details.} This is the first tight equivalence of this form in time-bounded Kolmogorov complexity that does not rely on an unproven assumption.

As a more sophisticated application of \Cref{thm:pkt_coding}, we establish an unconditional form of the main theorem from Antunes and Fortnow \cite{DBLP:conf/coco/AntunesF09}, which provides a characterization of the worst-case running times of languages that are in average polynomial-time over all $\mathsf{P}$-samplable distributions.  

We recall the following standard notion from average-case complexity (see, e.g.,~\cite{DBLP:journals/fttcs/BogdanovT06}). For an algorithm $A$ that runs in time $T_A\colon\bool^*\to\mathbb{N}$ and for a distribution $\caD$ supported over $\{0,1\}^*$, we say that $A$ runs in polynomial-time on average with respect to $\caD$ if there is some constant $\varepsilon>0$ such that
\[
\Exp_{x\sim \caD}\left[\frac{T_A(x)^{\varepsilon}}{|x|}\right]<1.
\]
As usual, we say that a distribution $\mathcal{D}$ is $\mathsf{P}$-samplable if it can be sampled in polynomial time.

	\begin{theorem}\label{thm:AF-pKt-intro}
    The following conditions are equivalent for any language $L \subseteq \{0,1\}^*$.
		\begin{itemize}
			\item[\emph{(}i\emph{)}] For every ${\sf P}$-samplable distribution $\mathcal{D}$, $L$ can be solved in polynomial-time on average with respect to $\mathcal{D}$.
			\item[\emph{(}ii\emph{)}] For every polynomial $p$, there exists a constant $b>0$ such that the running time of some algorithm that computes $L$ is bounded by $2^{O\left(\mathsf{\pK}^{p}(x)-\mathsf{K}(x)+b\log(|x|)\right)}$ for every input $x$.
		\end{itemize}
	\end{theorem}

In contrast, \cite{DBLP:conf/coco/AntunesF09} shows a \emph{conditional} characterisation result that employs $\K^t$ complexity in the expression that appears in Item (\emph{ii}).

\subsection{Techniques}\label{sec:techniques}

In this section, we provide an informal overview of our proofs and techniques.\\

\noindent \textbf{Efficient Coding Theorem for $\rKt$ (\Cref{thm:rkt_coding}).} 
Breaking down the result into its components, \Cref{thm:rkt_coding} shows that for any polynomial-time sampler $A$, there exist a probabilistic  polynomial-time algorithm $\comp$ and an algorithm $\dcomp$ with the following properties: $\comp$ on input an $n$-bit string $x$ and $\delta$ (which estimates from below the probability with which $A$ samples $x$), returns a codeword $c_x$ of length $\log(1/\delta) + \poly(\log n)$ such that $\dcomp$ with probability $\ge 0.99$ reconstructs $x$ in time $1/\delta \cdot \exp(\poly(\log n))$. Note that the probabilistic representation of $x$ certifying the $\rKt$ bound in \Cref{thm:rkt_coding} can be obtained from the codeword $c_x$ and $\dcomp$, and that obtaining a running time with a factor of $(1/\delta)^{1 + o(1)}$  is crucial in order to get a final $\rKt$ bound of the form $(2 + o(1)) \cdot \log(1/\delta)$. (Actually, $\comp$ does not have to depend on $A$, the $0.99$ can be $1-\varepsilon$ for arbitrary $\varepsilon >0$, and the $\poly(\log n)$ term is $O(\log n + \log^2 \log(1/\delta))$, but we omit these details in our discussion). We explain what are the challenges in obtaining $\comp$ and $\dcomp$ and how they are overcome. We remark that the construction is different from the approaches described in \cite{DBLP:conf/icalp/LuO21}.

$\dcomp$ can run the sampler $K := O(1/\delta)$ times and obtain a list of elements $S^*$ (the list of suspects) that with high probability contains $x$. $\comp$ has to provide information that allows $\dcomp$ to prune $S^*$ and find $x$. Since the algorithms do not share randomness, $\comp$ does not know $S^*$, and so compression has to work for any $S \subseteq \zo^n$ of size $K$, only assuming that $x \in S$. $\comp$ can use a  bipartite lossless expander graph $G$, which is a graph with the property that any set $S$ of left nodes with size $|S| \le K$ has at least $(1-\varepsilon)D |S|$ neighbors, where $D$ is the left degree. Such graphs are called $((1-\varepsilon)D, K)$ lossless expanders and they have numerous applications (see e.g., ~\cite{cap-rei-vad-wig:c:conductors,hoo-lin-wig:j:expander}). An extension of Hall's matching theorem shows that for any set $S$ of   $K$ left nodes, there is a matching that assigns to each $x \in S$, $(1-\varepsilon)D$ of its neighbors, so that no right node is assigned twice (i.e., the matching defines a subgraph with no collisions). $\comp$ can just pick the codeword $c_x$ to be one random neighbor of $x$. Then, $\dcomp$ can do the pruning of $S^*$ as follows. Having $S^*$ and $c_x$, it does the matching,  and,  since with probability $1-\varepsilon$,  $c_x$ is only assigned to $x$, $\dcomp$ can find $x$. There is one problem though. The  algorithms for maximum matching in general bipartite graphs  
do not run in linear time (see~\cite{DBLP:journals/corr/bestmaxflow, DBLP:conf/focs/Madry13}, and the references therein).  Therefore, the decompression time would have a dependency on $\delta$, which is
too large for us. Fortunately, lossless expanders can be used to do  ``almost'' matching faster. \cite{DBLP:journals/corr/abs-1911-04268} introduces \emph{invertible functions} (see~\Cref{d:invertible}) for the  more demanding task in which the elements of $S$ appear one-by-one and the matching has to be done in the online manner. We do not need online matching, but we take advantage of the construction in~\cite{DBLP:journals/corr/abs-1911-04268} to obtain a fast matching algorithm. It follows from~\cite{DBLP:journals/corr/abs-1911-04268}, that in a lossless expander it is possible to do a greedy-type of ``almost'' matching, which means that every left node in $S$ is matched to $(1-\varepsilon)D$ of its neighbors (exactly what we need), but with $\poly(\log n)$ collisions. The collisions can be eliminated with some additional standard hashing (see the discussion on page~\pageref{p:runtime} for details). As we explain on page~\pageref{p:runtime}, this leads to decompression time $K \cdot D \cdot \poly(n)$ and the length of the codeword $c_x$ is $\log |R| + |\text{hash-code}|$, where $R$ is the right set of the lossless expander. To obtain our result, the degree $D$ has to be $2^{\poly(\log n)}$ and $|R|$ has to be $K \cdot 2^{\poly(\log n)}$.  

Building on results and techniques from~\cite{DBLP:journals/jacm/GuruswamiUV09}, \cite{DBLP:journals/corr/abs-1911-04268} constructs a $((1-\varepsilon)D, K)$ explicit lossless expander with left side  $\zo^n$,  degree $D = 2^d$ for $d = O(\log(n/\varepsilon)\cdot \log k)$, and right side $R$, with size verifying $\log |R| = k + \log(n/\varepsilon) \cdot \log k$
(where $k:= \log K)$. To obtain in~\Cref{thm:rkt_coding} the dependency on $n$ to be $O(\log n)$ (which is optimal up to the constant in $O(\cdot)$), we show the existence of a $((1-\varepsilon)D, K)$ explicit expander with $d = O(\log n + \log (k/\varepsilon) \cdot \log k)$ and $\log |R| = k + O(\log n + \log(k/\varepsilon) \cdot \log k)$. This lossless expander is constructed by a simple composition of the above lossless expander from~\cite{DBLP:journals/corr/abs-1911-04268} with a lossless expander from~\cite{DBLP:journals/jacm/GuruswamiUV09},  with an appropriate choice of parameters (see~\Cref{s:betterprecision}).\\


\noindent \textbf{Conditional Lower Bound for Efficient Coding Theorems (\Cref{thm:coding_lower_bound}).} Our goal is to show that there is no \emph{efficient} coding theorem for $\rKt$ that achieves bounds of the form $(1 + \gamma - o(1)) \cdot \log(1/\delta) + \poly(\log n)$, under the assumption that \CryptoETH~holds for $\gamma \in (0,1)$. We build on an idea attributed to L.~Levin (see e.g.~\cite[Section 5.3]{TroyLeeThesis}). To provide an overview of the argument, let $G_n \colon \{0,1\}^{\ell(n)} \to \{0,1\}^n$ be a cryptographic generator of seed length $\ell(n) = n/2$ witnessing that \CryptoETH~holds. In other words, $G_n$ has security $2^{\gamma \cdot \ell(n)}$ against uniform adversaries. We define a sampler $S_n$ as follows. On input $x \in \{0,1\}^n$, which we interpret as a random string, it outputs $G_n(x')$, where $x'$ is the prefix of $x$ of length $\ell(n)$. We argue that if an \emph{efficient} algorithm $F$ is able to compress every string $y$ in the support of $\mathsf{Dist}(S_n)$, the distribution induced by the sampler $S_n$, to an $\rKt$ encoding of complexity $(1 + \gamma -\varepsilon) \cdot \log(1/\delta'(y)) + C \cdot (\log n)^C$, where $\delta'(y)$ is a lower bound on $\delta(y)$ (the probability of $y$ under $\mathsf{Dist}(S_n)$), we can use $F$ to break $G_n$. (Note that $F$ expects as input $n$, $y$, $\delta'$, and $\mathsf{code}(S)$.) 
		
		The (uniform) distinguisher $D$ computes roughly as follows. Given a string $z \in \{0,1\}^n$, which might come from the uniform distribution $U_n$ or from $G_n(U_{\ell(n)}) \equiv \mathsf{Dist}(S_n)$, $D$ attempts to use $F$ to \emph{compress} $z$ to a ``succinct'' representation, then checks if the computed representation \emph{decompresses} to the original string $z$. If this is the case, it outputs $1$, otherwise it outputs $0$. (Note that we haven't specified what ``succinct'' means, and it is also not immediately clear how to run $F$, since it assumes knowledge of a probability bound $\delta'$. For simplicity of the exposition, we omit this point here.) We need to argue that a test of this form can be implemented in time $2^{\gamma \cdot \ell(n)}$, and that it distinguishes the output of $G$ from a random string. 
		
		To achieve these goals, first note that a typical random string cannot be compressed to representations of length, say, $n - \mathsf{poly}(\log n)$, even in the much stronger sense of (time-unbounded) Kolmogorov complexity. Therefore, with some flexibility with respect to our threshold for succinctness, the proposed distinguisher is likely to output $0$ on a random string. On the other hand, if $F$ implements an efficient coding theorem that achieves $\rKt$ encodings of complexity $(1 + \gamma -\varepsilon) \cdot \log(1/\delta'(y)) + \mathsf{poly}(\log n)$, the following must be true. Using that the expected \emph{encoding length} of \emph{any} (prefix-free) encoding scheme is at least $H(\mathsf{Dist}(S_n))$, where $\mathsf{Dist}(S_n)$ is the distribution of strings sampled by $S_n$ and $H$ is the entropy function, we get (via a slightly stronger version of this result) that a non-trivial measure of strings $y$ in the support of $\mathsf{Dist}(S_n)$ have $\rKt$ \emph{encoding length} at least $(1 - \varepsilon/4) \cdot \log(1/\delta(y))$. Consequently, for such strings, an upper bound on $\rKt$ complexity of $(1 + \gamma -\varepsilon) \cdot \log(1/\delta'(y)) + \mathsf{poly}(\log n)$ when $\delta'(y)$ is sufficiently close to $\delta(y)$ implies that the running time $t$ of the underlying machine satisfies $\log t \leq (\gamma - \varepsilon/2)\log(1/\delta(y)) + \mathsf{poly}(\log n)$. Using that $\ell(n) = n/2$ and $\delta(y) \geq 2^{-\ell(n)}$ for any string $y$ in the support of $\mathsf{Dist}(S_n)$, it is easy to check that (asymptotically) $t \leq 2^{(\gamma - \varepsilon/4) \cdot \ell(n)}$. For this reason, we can implement a (slightly modified) distinguisher $D$ in time less than $2^{\gamma \cdot \ell(n)}$, by trying different approximations $\delta'(z)$ for an input string $z$ and by running the decompressor on the produced representation for at most $t$ steps on each guess for $\delta(z)$. By our previous discussion, a non-trivial measure of strings from $\mathsf{Dist}(S_n)$ will be accepted by $D$, while only a negligible fraction of the set of all strings (corresponding to the random case) will be accepted by $D$.
		
		Implementing this strategy turns out to be more subtle than this. This happens because $F$ is a \emph{probabilistic} algorithm which does not need to commit to a \emph{fixed} succinct encoding. We refer to the formal presentation in Section \ref{sec:cond_lower_bound} for details, where we also discuss the bound on the seed length $\ell(n)$.\\

    \noindent \textbf{Coding Theorem for \texorpdfstring{$\pK^t$}{pKt} (\Cref{thm:pkt_coding}) and Unconditional \cite{DBLP:conf/coco/AntunesF09} (\Cref{thm:AF-pKt-intro}).} The proof of our optimal coding theorem for $\pK^t$ builds on that of the \emph{conditional} coding theorem for $\K^t$ from \cite{DBLP:conf/coco/AntunesF09}, which can be viewed as a two-step argument. Roughly speaking, the first step is to show that if there is a polynomial-time sampler that outputs a string $x\in\bool^n$ with probability $\delta$, then the polynomial-time-bounded Kolmogorov complexity of $x$ is about $\log(1/\delta)+O(\log n)$ \emph{if we are given a random string}. After this, they ``derandomize'' the use of random strings using a certain pseudorandom generator, which exists under a strong derandomization assumption. Our key observation is that the use of random strings arises naturally in probabilistic Kolmogorov complexity, and particularly in this case the random strings can be ``embedded'' into the definition of $\pK^t$. As a result, we don't need to perform the afterward derandomization as in original proof of \cite{DBLP:conf/coco/AntunesF09}, and hence get rid of the derandomization assumption.

	Next, we describe how to use \Cref{thm:pkt_coding}, together with other useful properties of $\pK^t$, to obtain an unconditional version of Antunes and Fortnow's main result. Let $\mu$ be a Kolmogorov complexity measure, such as $\K^{\sf poly}$, $\rK^{\sf poly}$ or $\pK^{\sf poly}$. The key notion in the proof is the distribution (in fact, a class of semi-distributions) called $m_{\mu}$, which is defined as $m_\mu(x)\vcentcolon= 1/2^{\mu(x)}$. More specifically, following \cite{DBLP:conf/coco/AntunesF09}, it is not hard to show that, for every language $L$, $L$ can be decided in polynomial-time on average with respect to $m_\mu$ if and only if its worst-case running time is $2^{O(\mu(x)-\mathsf{K}(x))}$ on input $x$ (see \Cref{l:average_on_m}). Then, essentially, to show our result we argue that $L$ can be decided in polynomial time on average with respect to $m_{\mu}$ if and only if the same holds with respect to all $\mathsf{P}$-samplable distributions. 
	
	Recall that if a distribution $\caD$ \emph{dominates} another distribution $\caD'$ (i.e., $\mathcal{\caD}(x) \gtrsim \mathcal{\caD}'(x)$ for all $x$) and $L$ is polynomial-time on average with respect to $\caD$, then the same holds with respect to $\caD'$ (see \Cref{def:donmination} and \Cref{fact:domination_implies_average}). Therefore, to replace $m_\mu$ above with $\mathsf{P}$-samplable distributions, it suffices to show that $m_\mu$ is ``universal'' with respect to the class of $\mathsf{P}$-samplable distributions, in the following sense.
	\begin{enumerate}
		\item $m_\mu$ dominates \emph{every} $\mathsf{P}$-samplable distribution. (This is essentially an optimal source coding theorem for the Kolmogorov measure $\mu$.)
		\item $m_\mu$ is dominated by \emph{some} $\mathsf{P}$-samplable distribution.
	\end{enumerate}
	The above two conditions require two properties of the Kolmogorov measure $\mu$ that are somewhat conflicting: the first condition requires the notion of $\mu$ to be \emph{general} enough so that $m_\mu$ can ``simulate'' \emph{every} $\mathsf{P}$-samplable distribution, while the second condition needs $\mu$ to be \emph{restricted} enough so that $m_\mu$ can be ``simulated'' by \emph{some}  $\mathsf{P}$-samplable (i.e., simple) distribution. For example, if $\mu$ is simply the time-unbounded Kolmogorov complexity $\K$ (or even the polynomial-space-bounded variant), then it is easy to establish an optimal source coding theorem for such a general Kolmogorov measure; however it is unclear how to sample in polynomial-time a string $x$ with probability about $1/2^{\K(x)}$, so in this case $\mu$ does not satisfy the second condition. On the other hand, if $\mu$ is some restricted notion of time-bounded Kolmogrov complexity measure such as $\K^{\sf poly}$ or $\rK^{\sf poly}$, then one can obtain polynomial-time samplers that sample $x$ with probability about $1/2^{\K^{\poly}(x)}$ or $1/2^{\K^{\poly}(x)}$ (up to a polynomial factor); however, as in \cite{DBLP:conf/coco/AntunesF09}, we only know how to show an \emph{optimal} source coding theorem for $\K^{\sf poly}$ (or $\rK^{\sf poly}$) under a derandomization assumption. Therefore, in this case $\mu$ does not satisfies the first condition. Our key observation is that the notion $\pK^{\sf poly}$, which sits in between $\K$ and $\K^{\poly}$ (or $\rK^{\poly}$),\footnote{We can show that for every $x\in\bool^*$ and every computable time bound $t\colon\mathbb{N}\to\mathbb{N}$, $\K(x)\lesssim\pK^{t}(x)\leq \rK^t(x)\leq\K^t(x)$.} satisfies both conditions described above (see \Cref{l:main,l:P_samplable_dominates_m}).

	\section{Preliminaries}\label{sec:preliminaries}

\noindent \textbf{Time-bounded Kolmogorov complexity.} For a function $t \colon \mathbb{N} \to \mathbb{N}$, a string $x$, and a universal Turing machine $U$, let the time-bounded Kolmogorov complexity be defined as
  \[
    \K_U^t(x) = \min_{p \in \{0,1\}^*} \left\{|p| \mid U(p)~\textnormal{outputs $x$ in at most $t(|x|)$ steps} \right\}.
  \]
  A machine $U$ is said to be {\em time-optimal} if for every machine $M$ there exists a constant $c$ such that for all $x \in \{0,1\}^n$ and $t \colon \mathbb{N} \to \mathbb{N}$ satisfying $t(n) \geq n$,
  \[
    \K^{ct\log t}_U(x) \le \K^{t}_M(x) + c,
  \]
where for simplicity we write $t = t(n)$. It is well known that there exist time-optimal machines~~\cite[Th. 7.1.1]{li-vit:b:kolmbook-four}. In this paper, we fix such a machine~$U$, and drop the index $U$ when referring to time-bounded Kolmogorov complexity measures.	It is also possible to consider prefix-free notions of Kolmogorov complexity. However, since all our results hold up to additive $O(\log |x|)$ terms, we will not make an explicit distinction.

	 Henceforth we will not distinguish between a Turing machine $\mathcal{M}$ and its encoding $p$ according to $U$. If $p$ is a probabilistic Turing machine, we define $t_p \in \mathbb{N} \cup \{\infty\}$ to be the maximum number steps it takes  $p$ to halt  on input   $\lambda$ (the empty string), where the maximum is over all branches of the probabilistic computation.\\
	
	\noindent \textbf{$\rKt$ complexity and probabilistic representations.} A probabilistic representation of a string $x$ is a probabilistic Turing machine $p$ that on input $\lambda$ halts with $x$ on the output tape with probability at least $2/3$. The $\rKt$-complexity of a string $x$ is the minimum, over all probabilistic representations $p$ of $x$,  of $p + \log t_p$. A \emph{probabilistic representation} $p$ of $x$ \emph{certifies $\rKt$-complexity bounded by $\Gamma$} if $|p|+ \log t_p \le \Gamma$.\\

	\noindent \textbf{Distributions and semi-distributions.} We consider distributions over the set $\bool^*$. We will identify a distribution with its underlying probability density function of the form $\caD\colon \bool^{*}\to[0,1]$.
	A distribution $\caD$ is a \emph{semi-distribution} if $\sum_{x\in \bool^*}\caD(x)\leq 1$, and is simply called a \emph{distribution} if the sum is exactly $1$. In this subsection and \Cref{sec:AF}, we will use the word ``distribution'' to refer to both distribution and semi-distribution.\\
	
		\noindent \textbf{Samplers.} A \emph{sampler} is a probabilistic algorithm $A$ with inputs in $\{1\}^n$  such that $A(1^n)$ outputs a string $x \in \zo^n$.\footnote{For simplicity, we assume that $A(1^n)$ samples a string of length $n$.  Our coding theorems also hold  for algorithms used to define $\P$-samplable distributions, see~\Cref{d:psamplabled}, with obvious changes in the proofs. Also, as in \cite{DBLP:conf/icalp/LuO21}, our results can be easily generalised to samplers that on $1^n$ output strings of arbitrary length. In this case, while the length of $x$ might be significantly smaller than $n$, an additive overhead of $\log n + O(1)$ is necessary in our coding theorems, as we need to encode $1^n$.} It defines a family of distributions $\{\mu_{A,n}\}_{n \in \mathbb{N}}$, where $\mu_{A,n}$ is the distribution on $\zo^n$ defined by $\mu_{A,n}(x) = \Prob_A[A(1^n)=x]$.\\
		
		\noindent \textbf{Average-case complexity.} We now review some standard definitions and facts from average-case complexity. We refer to the survey \cite{DBLP:journals/fttcs/BogdanovT06} for more details.

	\begin{definition}[Polynomial-time Samplable \cite{DBLP:journals/jcss/Ben-DavidCGL92}]\label{d:psamplabled}
		A distribution $\caD$ is called $\P$-samplable if there exists a polynomial $p$ and a probabilistic algorithm $M$ such that for every $x\in\bool^*$, $M$ outputs $x$ with probability $\caD(x)$ within $p(|x|)$ steps.
	\end{definition}

	\begin{definition}[Polynomial Time on Average \cite{DBLP:journals/siamcomp/Levin86}]
		Let $A$ be an algorithm and $\caD$ be a distribution. We say that $A$ runs in polynomial-time on average with respect to $\caD$ if there exist constants $\varepsilon$ and $c$ such that,
		\[
		    \sum_{x\in\bool^*} \frac{t_A(x)^{\varepsilon}}{|x|} \caD(x) \leq c,
		\]
		where $t_A(x)$ denotes the running time of $A$ on input $x$. For a language $L$ we say that $L$ can be solved in polynomial time on average with respect to $\caD$ if there is an algorithm that computes $L$ and runs in polynomial-time on average with respect to $\caD$.
	\end{definition}	
	\begin{definition}[Domination]\label{def:donmination}
		Let $\caD$ and $\caD'$ be two distributions. We say that $\caD$ \emph{dominates} $\caD'$ if there is a constant $c>0$ such that for every $x\in\bool^*$,
		\[
		\caD(x)\geq \frac{\caD'(x)}{|x|^c}.
		\]
	\end{definition}
	
	\begin{fact}[See e.g., {\cite[Lemma 3.3]{DBLP:conf/coco/AntunesF09}}]\label{fact:domination_implies_average}
		Let $\caD,\caD'$ be two distributions, and let $A$ be an algorithm. If 
		\begin{itemize}
			\item $A$ runs in polynomial time on average with respect to $\caD$, and
			\item $\caD$ dominates $\caD'$
		\end{itemize}
		Then $A$ also runs in polynomial time on average with respect to $\caD'$.
	\end{fact}

	\section{Coding Theorems for \texorpdfstring{$\mathsf{rKt}$}{rKt} Complexity}
	
\subsection{Efficient Coding Theorem with Tighter Parameters}\label{sec:efficient_rkt_coding}

	We prove the result stated  in~\Cref{thm:rkt_coding}.  It involves a function 
	\begin{equation}\label{e:alpha}
	\alpha(n,1/\delta, \varepsilon) = O( \log n + (\log \log 1/\delta + \log(1/\varepsilon)) \cdot \log\log 1/\delta \,),
	\end{equation}
	which bounds the additive precision  term in the length of codewords. The constant hidden in $O(\cdot)$ is derived from the proof of~\Cref{t:invertible} (stated below).  The key fact is stated in the following result.

	\begin{theorem}\label{t:compsamplable}
		There exist a probabilistic polynomial-time algorithm $\comp$ and a probabilistic algorithm $\dcomp$ such that for every $n$-bit string $x$, and every rationals $\delta > 0$ and $\varepsilon > 0$,\\
		
		\vspace{-0.3cm}
		
		$\bullet$ $\comp$ on input $x, 1/\delta, \varepsilon$ outputs a string $p$ that has with probability $1$ length $\log(1/\delta) + \alpha(n,1/\delta, \varepsilon)$, and
		\smallskip
		
		$\bullet$ If $x$ can be sampled by a polynomial-time sampler $A$ with probability at least $\delta$, then,  with probability at least $1-\varepsilon$, $\dcomp$ on input $A$ and $p$ outputs $x$. Moreover,  $\dcomp$ on input $A$ and  $p$ halts in $t_D := 1/\delta \cdot 2^{O(\alpha(n, 1/\delta, \varepsilon))}$ steps with probability $1$. The constant in $O(\cdot)$ depends on $A$, and the two probabilities are over the randomness of $\comp$ and the randomness of $\dcomp$.
	\end{theorem}
		\begin{proof}
The proof uses the \emph{invertible functions} from~\cite{DBLP:journals/corr/abs-1911-04268}. A $(k, \varepsilon)$-invertible function is a probabilistic function that on input $x$ produces a random fingerprint of $x$. The invertibility property requires that there exists a deterministic algorithm that on input a random fingerprint of $x$ and a set $S$ (the ``list of suspects'')  of size at most $2^k$ that contains $x$, with probability $1-\varepsilon$ identifies $x$ among the suspects. 
\begin{definition}\label{d:invertible}
 A function $F: \zo^n \times \zo^d \mapping \zo^{k+\Delta}$ is $(k, \varepsilon)$-invertible if there exists a partial  function $g$ 
  mapping a set $S$ of $n$-bit strings and a $(k+\Delta)$-bit string $y$ into $g_S(y) \in \zo^{n}$ 
  such that for every set $S$ of size at most $2^k$ strings and every $x$ in $S$
\begin{equation}\label{e:inv}
\Prob_\rho[g_S(F(x, \rho)) = x] \geq 1- \varepsilon.
\end{equation}

\end{definition}

The actual $(k,\varepsilon)$-invertible function that we use is given in the following theorem, which is essentially Theorem 2.1  in~\cite{DBLP:journals/corr/abs-1911-04268}, with an improvement of the precision term. (The result in~\cite{DBLP:journals/corr/abs-1911-04268} is stronger, because the ``list of suspects'' $S$ can be presented to the inverter $g$ from~\Cref{e:inv} in an \emph{online} manner, but in our application we do not need the online feature.)
\begin{theorem}\label{t:invertible}
  There exists  a probabilistic algorithm $\ff$ that on input $\varepsilon>0$, $k$ and string $x$, uses $d = \alpha(|x|, 2^k, \varepsilon)$ random bits and outputs  
  in time polynomial in~$|x|$ a string $\ff_{\varepsilon,k}(x)$ of length $k+ \alpha(|x|, 2^k, \varepsilon)$, 
  such that for all $\varepsilon > 0$ and $k$, the function $x \mapsto \ff_{\varepsilon,k}(x)$ is 
  $(k,\varepsilon)$-invertible. Moreover, the inverter $g$ that satisfies~\Cref{e:inv} runs in time $|S| \cdot 2^d \cdot \poly(|x|)$.
\end{theorem}

\label{p:runtime}  This theorem is proven in~\cite{DBLP:journals/corr/abs-1911-04268} (for a somewhat larger $d$, but this does not affect the arguments), however the time bound is not explicitly stated. Therefore, we  describe below the invertible  function and verify the bound. We also need to  explain how to obtain the value $d=\alpha(|x|, 2^k, \varepsilon) = O(\log n + \log (k/\varepsilon) \cdot \log k)$ claimed in~\Cref{t:invertible} (Theorem 2.1  in~\cite{DBLP:journals/corr/abs-1911-04268} has $d= O(\log (n/\varepsilon) \cdot \log k)$). This is done in~\Cref{s:betterprecision}.
\smallskip

The function $F(x,\rho)$ (the random fingerprint of $x$) is formed by concatenating 2 strings, $F_1(x, \rho_1)$ and $F_2(x, \rho_2)$ (here, $\rho=(\rho_1, \rho_2)$ is the randomness of $F$). The first one is obtained by evaluating an explicit conductor (see further), and the second one is a standard hash code. The reconstruction of $x$ from the fingerprint is done in two pruning stages. 
In the first stage,  the conductor part will be used to reduce the list of suspects $S$ (which includes $x$) of size  $2^k$ to a list $\tilde{S}$ of size $2^{\alpha(n,2^k,\eps)}$ (which also includes $x$ w.h.p), and in the second stage, the second component of the fingerprint is used  to select one string in $\tilde{S}$, which with probability $\geq 1-\varepsilon$ is $x$. 
The second reduction is simple: a greedy algorithm is used that selects the first string in $\tilde{S}$ for which the hash code matches. To distinguish a string $x$ from $s$ other strings in this way, one can use a hash code with prime numbers, which has size $2\log s + O(\log n)$. 

The non-trivial part is the first stage, which reduces the list $S$ of size $2^k$ to a list of quasi polynomial size. (A greedy algorithm would require hash codes of bitlength $2k$ instead of~$k$.) Let $\mcX = \{0,1\}^n$ and $\mcY = \{0,1\}^k$. A hash code $H : \mcX \times \{0,1\}^r \rightarrow \mcY$ defines a bipartite graph with left set $\mcX$, right set $\mcY$, and left degree $2^r$. 
(We will use as the bipartite graph an explicit conductor graph with $r \le \alpha(n, 1/\delta, \varepsilon)$, this is the above $F_1(x, \rho_1)$.)

We explain now the first reduction of the list of suspects. Given a set $S$ and a right node $y \in \mcY$ (which in our application is $F_1(x,\rho_1)$), the algorithm will output at most $(1+\log |S|) 2^{r+1}$ strings from $S$ as follows. It iterates through all elements in $S$ and selects for the output list the first $2^{r+1}$ left neighbors of $y$. If there are no more such neighbors, the algorithm is finished. Otherwise, it will compute the set $S'$ of all left nodes for which more than a fraction $2\eps$ of right nodes have more than $2^{r+1}$ collisions. Then it will run the reduction algorithm recursively on $S'$ and append the output list of the recursive call. 

This is applied to a function $H$ with $r \le \alpha(n, 1/\delta, \varepsilon)$ that satisfies a conductor property. More precisely, we need the properties of a lossless conductor, an object which is essentially equivalent to a bipartite lossless expander, that we used in the high-level description in~\Cref{sec:techniques}. This property implies that $|S'| \le |S|/2$, and hence, by induction, we obtain the bound $(1+\log |S|)2^{r+1}$ for the size of the output list, which is passed to the second reduction.

Let us now evaluate the runtime. In each recursive call we iterate over all elements in $S$ and compute the list of right neighbors, which takes time $|S|\cdot 2^r \cdot \poly(|x|)$. To calculate $S'$, one maintains a counter for each right node.  Then we iterate through all elements of $S$ and for each element, increment the counters of its right neighbors by one. Then we collect all right nodes with counters that exceed $2^r$ into the set $S'$ and start the recursion. Thus, since a left node has $2^r$ neighbors, one recursive call takes $|S|\cdot 2^r \cdot \poly(|x|)$ steps. There may be $\log |S|$ recursive calls, but $\log|S| \le |x|$. Thus, the total time for the first reduction is bounded by $|S|\cdot 2^r \cdot \poly(|x|) \leq  |S|\cdot 2^d \cdot \poly(|x|)$. The second reduction, using standard hashing, takes time bounded by $2^d \cdot \poly(|x|)$. Thus, we got the claimed runtime.

\if01
\bnote{explanations time bound}
This theorem is proven in~\cite{DBLP:journals/corr/abs-1911-04268} but the time bound is not explicitly stated. Therefore, we first describe the inverter function and verify the bound. 

The invertible is formed by concatenating 2 strings. The first is obtained by evaluating an explicit conductor, (see further), and the second is a standard hash code. The conductor part will be used to reduce the list of suspects $S$ of size  $2^k$ to a list $\tilde{S}$ of size $2^{\alpha(n,k,\eps)}$, and the second one will select one string in $\tilde{S}$. For the second, a greedy algorithm is used that selects the first strings in $\tilde{S}$ for which the hash code matches. To distinguish a string $x$ from $s$ other strings in this way, one can use a hash code with prime numbers, which has size $2\log s + O(\log n)$. 

The non-trivial part is to reduce the list $S$ of size $2^k$ to a list of quasi polynomial size. (A greedy algorithm would require hash codes of bitlength $2k$ instead of~$k$.) Let $\mcX = \{0,1\}^n$ and $\mcY = \{0,1\}^k$. A hash code $H : \mcX \times \{0,1\}^r \rightarrow \mcY$ defines a bipartite graph with left set $\mcX$, right set $\mcY$, and left degree $2^r$. 
(We will us an explicit conductor graph with $r \le \alpha(n, 1/\delta, \varepsilon)$.)

We explain the reduction of the list of suspects. Given a set $S$ and a right node $y \in \mcY$ the algorithm will output at most $(1+\log |S|) 2^{r+1}$ strings from $S$ as follows. It iterates through all elements in $S$ and selects the first $2^{r+1}$ left neighbors of $y$. If there exists no more such neighbors, the algorithm is finished. Otherwise, it will compute the set $S'$ of all left nodes for which more than a fraction $2\eps$ of right nodes have more than $2^{r+1}$ collisions. Then it will run the selection algorithm recursively on $S'$ and append the output of the recursive call. 

This is applied to a function $H$ with $r \le \alpha(n, 1/\delta, \varepsilon)$ that satisfies a conductor property. This property implies that $|S'| \le |S|/2$, and hence, the bound $(1+\log |S|)2^{r+1}$ follows by induction.

We now understand why the run-time is $|S|\cdot 2^d \cdot \poly(|x|)$. In each recursive call we maintain a counter for each right node. Then we iterate through all elements of $S$ and for each element, increment the counters of its right neighbors by one. Then we collect all right nodes with counters that exceed $2^r$ into the set $S'$ and start the recursion.
\bnote{end of new text}

We explain how to obtain~\Cref{t:invertible} from Theorem 2.1  in~\cite{DBLP:journals/corr/abs-1911-04268} in~\Cref{s:betterprecision}.
\smallskip
\fi

We now define $\comp$ and $\dcomp$ with the properties required in~\Cref{t:compsamplable}. We assume that $\delta \geq 2^{-|x|}$, because otherwise the trivial compressor, that compresses $x$ to $x$ itself, satisfies the conditions.

The function $\comp$ on input $x, 1/\delta, \varepsilon$, takes $K := \lceil \ln (1/\varepsilon) \cdot  (1/\delta) \rceil$ and runs $\ff$ from~\Cref{t:invertible} on input $x, \log K, \varepsilon$,  which produces a string $p$ of length 
$\log K  + \alpha(|x|, K, \varepsilon)$. We can assume that $K$ is a power of $2$ (otherwise we replace it in the following arguments with the smallest power of $2$ larger than it), and thus it can be described with $\log \log K$ bits.
$\comp$ also appends  $|x|$ and the short description of $K$ to $p$ encoded in a self-delimited way. This takes only an additional $O(\log|x| + \log \log K)$ bits. By scaling up the constant in the definition of $\alpha(\cdot, \cdot, \cdot)$, we get that the length of $p$ is bounded by
\[
\log(1/\delta) + \alpha(|x|, K, \varepsilon).
\]

$\dcomp$ first produces a list of suspects $S$ by running the sampler $A$ on input $1^{|x|}$ (with $|x|$ extracted from $p$), $K$ times ($K$ is also extracted from $p$) and taking $S$ to be the set of samples that are obtained. We have $|S| \le  K$ and, since $x$ is sampled with probability at least $\delta$, the probability that $x$ is not in $S$ is bounded by $(1-\delta)^K < \varepsilon$.  This step takes time $K \cdot \poly(|x|)$, with the degree of the polynomial depending on $A$.

Next,  $\dcomp$ runs the inverter $g$ on input $S$ and $p$. From~\Cref{e:inv} we infer that if $x \in S$, then, with probability $1-\varepsilon$, this computation reconstructs $x$.
Overall, $\dcomp$ reconstructs $x$ with probability $1-2\varepsilon$, and by the ``Moreover ...'' part of~\Cref{t:invertible}, its runtime is 
\[
t_D := |S| \cdot 2^{d} \cdot \poly (|x|) + K \cdot \poly(|x|) = K \cdot 2^{\alpha(|x|, K, \varepsilon)} \cdot \poly(|x|) = (1/\delta) \cdot 2^{O(\alpha(|x|, 1/\delta, \varepsilon))}. 
\]
The conclusion follows after a rescaling of $\varepsilon$.
\end{proof}
	We now readily obtain the announced result. 
		\begin{corollary}[Efficient coding for $\rKt$] \label{t:effrKt}
	Let $x \in \{0,1\}^n$. Suppose there is a polynomial-time sampler $A$  such that $A(1^n)$ outputs $x$ with probability at least $\delta > 0$. Then, for every $\varepsilon >0$
\[
\rKt_\varepsilon(x) \;\leq\; 2 \log(1/\delta) + O(\log (\alpha(|x|, 1/\delta, \varepsilon)),
		\]
			where the constant hidden in $O(\cdot)$ depends on $A$. 

Moreover, there is a probabilistic polynomial time algorithm that
		on input $x, 1/\delta, \varepsilon$ and the code of $A$, outputs with probability $1-\varepsilon$, a probabilistic representation of $x$ certifying the above $\rKt$-complexity. 
	\end{corollary}
	\begin{proof}
		Indeed, in the proof of~\Cref{t:compsamplable}, we have seen that $\comp$, with probability $1-\varepsilon$, produces a string $p$, which, given the sampler, allows the reconstruction of $x$ with  probability at least $1-\varepsilon$. More precisely, for every list of suspects $S$ produced by $\dcomp$ with $x \in S$, a fraction of $(1-\varepsilon)$ of $p$'s allow $\dcomp$ to reconstruct $x$. With a standard averaging argument, we can change the order of quantifiers, and show that for a fraction of $1-\sqrt{\varepsilon}$ of $p$'s, it holds that for $1-\sqrt{\varepsilon}$ of $S$'s as above, $\dcomp$ can reconstruct $x$.\footnote{The averaging argument is done with respect to the distribution over sets $S$ induced by $\dcomp$.}  Since $\comp$ is a polynomial-time probabilistic algorithm and both the length of $p$ and the logarithm of the reconstruction time $t_D$  are bounded by $\log(1/\delta) + O(\alpha(|x|, 1/\delta, \varepsilon))$, the conclusion follows by scaling $\varepsilon$ and taking the probabilistic representation of $x$ to be $\langle p, \mathsf{code}(A) \rangle$, where $\langle \cdot, \cdot \rangle$ is some canonical self-delimiting pairing of strings.
	\if01
	Indeed, in the proof of~\Cref{t:compsamplable}, we have seen that $\comp$, with probability $1-\varepsilon$, produces a string $p$, which, given the sampler, allows the reconstruction of $x$ with  probability at least $1-\varepsilon$.  Since $\comp$ is a polynomial-time probabilistic algorithm and both the length of $p$ and the logarithm of the reconstruction time $t_D$  are bounded by $\log(1/\delta) + O(\alpha(|x|, 1/\delta, \varepsilon))$, the conclusion follows
	by taking the probabilistic representation of $x$ to be $\langle p, \mathsf{code}(A) \rangle$, where $\langle \cdot, \cdot \rangle$ is some canonical self-delimiting pairing of strings.
	\fi
	\end{proof}
	We can assume that $\log(1/\delta) \le n$, since otherwise, as we have already noted, $\comp$ can trivially simply return   $x$. Therefore, the overhead in~\Cref{t:effrKt} is bounded by the simpler  term $O(\log(n/\varepsilon) \cdot \log n)$.

		\subsubsection{Improving the Precision Term in the Invertible Function in~\texorpdfstring{\cite{DBLP:journals/corr/abs-1911-04268}}{[BZ19]}}\label{s:betterprecision}

In Theorem 2.1 in~\cite{DBLP:journals/corr/abs-1911-04268}, an invertible function is constructed, in which the precision term is $O(\log(n/\varepsilon) \cdot \log k)$. We need to improve it  to $O(\log n + (\log k/\varepsilon) \cdot \log k)$, to obtain the precision claimed in~\Cref{t:invertible}. The same idea is used in~\cite[Th. 4.21]{DBLP:journals/jacm/GuruswamiUV09}.

	We start with some standard concepts from the theory of pseudo-randomness (see~\cite{DBLP:journals/fttcs/Vadhan12}). A \emph{source} is a random variable whose realizations are binary strings. A source  has {\em min-entropy $t$} if each value has probability at most $2^{-t}$. The statistical distance between two measures $P$ and $Q$ with the same domain is $\sup |P(S)-Q(S)|$ (supremum is over all subsets $S$ of the common domain of $P$ and $Q$).
 Given a set $B$, we denote $U_B$  to be a random  variable that is uniformly distributed on $B$.

\begin{definition}[Condensers and lossless conductors]\label{d:conductor}~\\\vspace{-0.5cm}
  \begin{itemize}
      \item [\emph{(}a\emph{)}] A function $C: \zo^n \times \zo^d \mapping \zo^{m}$ is a $t \rightarrow_{\varepsilon} t'$ condenser, if for every $S \subseteq \zo^n$ of size at least $2^t$, the random variable $X= C\left(U_S, U_{\zo^d}\right)$ is $\varepsilon$-close to a random variable $\widetilde{X}$ that has min-entropy at least $t'$. 
      \item [\emph{(}b\emph{)}] A function $C: \zo^n \times \zo^d \mapping \zo^{m}$ is a $(\tm, \varepsilon)$ lossless conductor if it is a  
      $t \rightarrow_\varepsilon t+d$ condenser for all $t \le \tm$.
  \end{itemize}
\end{definition}
In the proof in~\cite{DBLP:journals/corr/abs-1911-04268}, the following lossless conductor is obtained, which is next composed with a certain hash function to obtain the invertible function.

\begin{theorem}[\cite{DBLP:journals/corr/abs-1911-04268}, implicit in Theorem 2.1]\label{t:bz}
For every $n$, $\tm \leq n$, $\varepsilon > 0$, there exists an explicit $(\tm, \varepsilon)$ lossless conductor $C_{\bz} : \zo^n \times \zo^{d_{\bz}} \mapping \zo^{m_{\bz}}$  with $d_{\bz} = O( \log (n/\varepsilon) \cdot \log (\tm)$ and $m_{\bz} \leq \tm + O( \log (n/\varepsilon) \cdot \log (\tm)$.
\end{theorem}

To obtain the desired improvement, we need to replace $C_{\bz}$ in the construction in Theorem 2.1 in~\cite{DBLP:journals/corr/abs-1911-04268} with a $(\tm, \varepsilon)$ lossless conductor $C$  with seed length $d_C =O((\log n + \log (\tm/\varepsilon)) \cdot  \log \tm)$ and output length $m_C \leq \tm + O((\log n + \log (\tm/\varepsilon)) \cdot  \log \tm)$.

This is obtained by composing $C_{\bz}$ with the following lossless conductor of Guruswami, Umans, and Vadhan.

\begin{theorem}[\cite{DBLP:journals/jacm/GuruswamiUV09}, Theorem 4.4]\label{t:guv}
For every $n$, $\tm \leq n$, $\varepsilon > 0$, there exists an explicit $(\tm, \varepsilon)$ lossless conductor $C_{\guv} : \zo^n \times \zo^{d_{\guv}} \mapping \zo^{m_{\guv}}$  with $d_{\guv} \leq \log n + \log (\tm) + \log(1/\varepsilon) +1$ and $m_{\guv} \leq d_{\guv} \cdot(\tm+2)$.
\end{theorem}

Now, as announced, we obtain the lossless conductor $C$, by composing $C_{\guv}$ and $C_{\bz}$. Namely, $C: \zo^n \times \zo^{d_{\guv}+ d_{\bz}} \mapping \zo^{m_{\bz}}$ is defined by \[
C(x, (y_1, y_2)) = C_{\bz}(C_{\guv} (x, y_1), y_2).
\]
Let us specify the parameters.  We first condense with $C_{\guv}$ for sources $X$ with min-entropy $t \leq \tm$  using a seed of length $d_{\guv} \leq \log n + \log (\tm) + \log(1/\varepsilon) +1$ and obtain an output $X_1$ of length $m_{\guv} \leq d_{\guv} \cdot(\tm+2)$ and min-entropy at least $t+d_{\guv}$. Then we further condense $X_1$ with $C_{\bz}$ with parameters $d_{\bz}$ and $m_{\bz}$ set-up for sources with input length $m_{\guv}$ and min-entropy bounded by $\tm' = \tm + d_{\guv}$. In this way, we obtain a $(\tm, 2\varepsilon)$ lossless conductor for all $t \leq \tm$, with the parameters $d_C$ and $m_C$ announced above, except for the case  $\tm = o(\log n)$. But for such small $\tm$, we can take the conductor that simply outputs the seed. 

Thus, we have obtained the following lossless conductor.
\begin{theorem}[The new conductor]\label{t:newconductor}
For every $n$, $\tm \leq n$, $\varepsilon > 0$, there exists an explicit $(\tm, \varepsilon)$ lossless conductor $C: \zo^n \times \zo^{d_C} \mapping \zo^{m_C}$  with $d_C = O( \log n + \log (\tm/\varepsilon) \cdot \log (\tm))$ and $m_{C} \leq \tm + O(\log n +  \log (\tm/\varepsilon) \cdot \log (\tm))$.
\end{theorem}

\begin{remark}
It has been observed in~\cite{DBLP:journals/combinatorica/Ta-ShmaUZ07} that lossless conductors are essentially equivalent to lossless bipartite expanders, which have numerous applications. Therefore the  conductor in~\Cref{t:newconductor} is of independent interest (it is already being used in a work in progress of some of the authors). The main lossless conductor (or, equivalently, lossless bipartite expander) of Guruswami, Umans, and Vadhan~\cite{DBLP:journals/jacm/GuruswamiUV09} (different from the one in~\Cref{t:guv}) has a better seed length of $d= (1+1/\alpha) (\log n + \log \tm + \log (1/\varepsilon)) + O(1)$, for any $\alpha \in (0,1)$, but the output length $m= (1+\alpha) \tm + 2d$ is larger, and would not have produced the optimal coding theorem.
\end{remark}

	\subsection{Existential Coding Theorem Under a Derandomization Assumption}\label{sec:existential_coding}

	The argument is a straightforward adaptation of a proof from \cite{DBLP:conf/coco/AntunesF09}, and we include a sketch here for completeness.

	\begin{theorem}
     Assume there is a language $L\in {\sf BPTIME}\left[2^{O(n)}\right]$ that requires nondeterministic circuits of size $2^{\Omega(n)}$ for all but finitely many $n$. Suppose there is an algorithm $A$ for sampling strings that runs in time $T(n)$ such that $A(1^n)$ outputs a string $x \in \{0,1\}^n$ with probability at least $\delta > 0$. Then
		\[
		\rKt(x) \,\leq\,  \log(1/\delta) +  O\!\left(\log T(n)\right),
		\]
		where the constant behind the $O(\cdot)$ depends on $|A|$ and is independent of the remaining parameters.
	\end{theorem}
	\begin{proof}[Proof Sketch]
	As briefly described in \Cref{sec:techniques}, the proof of the coding theorem in \cite{DBLP:conf/coco/AntunesF09} first shows (unconditionally) that if we are given a typical random string $r$ of length $\poly(T(n))$, then there exists some string $\alpha\in\bool^{\ell}$, where $\ell\leq \log(1/\delta) + O(1)$, from which one can recover the string $x$ in time $\poly(T(n))$. Roughly speaking, the string $r$ encodes a good \emph{hitting set generator} $H\colon\bool^{\ell}\to\bool^{T(n)}$, and $\alpha$ is an input to $H$ so that using $H(\alpha)$ as its internal randomness, $A(1^n)$ outputs $x$. Then it was observed in \cite{DBLP:conf/coco/AntunesF09} (attributed to van Melkebeek) that one can further encode such an $r$ using some string  $r_0\in\bool^{O(\log T(n))}$ if we have an optimal PRG $G\colon\bool^{O(\log T(n))}\to\bool^{|r|}$ that fools \emph{nondeterministic circuits}. That is, $G(r_0)=r$. Then given $r_0$, we can eventually obtain the string $r$ and use it along with $\alpha$ to recover $x$ \emph{deterministically} in time $\poly(T(n)$, which implies
	\[
	\Kt(x) \,\leq\,  \log(1/\delta) +  O\!\left(\log T(n)\right).
	\]
	The existence of such a PRG follows from the assumption that there exists a language $L\in {\sf DTIME}\!\left[2^{O(n)}\right]$ that requires nondeterministic circuits of size $2^{\Omega(n)}$ for all but finitely many $n$ \cite{DBLP:journals/jacm/ShaltielU05}. It turns out that if the language above is in ${\sf BPTIME}\!\left[2^{O(n)}\right]$ instead of ${\sf DTIME}\!\left[2^{O(n)}\right]$, we can still obtain an optimal \emph{pseudodeterministic} PRG; this again follows from the construction in \cite{DBLP:journals/jacm/ShaltielU05}. A PRG $G$ is pseudodeterministic if there is a randomized algorithm that, given a seed, computes the output of $G$ on this seed with high probability. That is, if we have such a PRG, then in the above argument, we can obtain $r=G(r_0)$ with high probability, which then allows us to recover $x$ in the same way but \emph{probabilistically}. Therefore, we get
	\[
		\rKt(x) \,\leq\,  \log(1/\delta) +  O\!\left(\log T(n)\right),
	\]
	as desired.
\end{proof}

		\section{Lower Bounds for Efficient Coding Theorems}\label{sec:cond_lower_bound}

	\subsection{Conditional Optimality of the Efficient Coding Theorem for \texorpdfstring{$\mathsf{rKt}$}{rKt}}\label{sec:cond_rKt}

	We introduce the following hypothesis, which postulates the existence of a cryptographic PRG $G$ of exponential security.

	\begin{hypothesis}[Cryptographic Exponential Time Hypotheses] For a constant $\gamma \in (0,1)$, we let \CryptoETH~be the following statement. There is a function family  $G = \{G_n\}_{n \geq 1}$, where $G_n \colon \{0,1\}^{\ell(n)} \to \{0,1\}^n$, such that the following holds:
		\begin{itemize}
		 \item The seed length $\ell(n)$ can be computed in time polynomial in $n$ and $(\log n)^{\omega(1)} \le \ell(n) \le n/2$. 
			\item $G$ is efficiently computable in the output length, i.e., there is a deterministic polynomial time algorithm $A$ that, given $1^n$ and an input $x \in \{0,1\}^{\ell(n)}$, outputs $G_n(x)$ in time $\poly(n)$,
			
			\item $G_n$ has security $2^{\gamma \cdot \ell(n)}$, i.e., for every probabilistic algorithm $D$ that runs in time $O(2^{\gamma \cdot \ell(n)})$ on inputs of length $n$, there exists $n_0 \in \mathbb{N}$ such that, for every $n \geq n_0$,
			\[
			\left | \;\Prob_{\bm{D},\, \bm{x} \sim \{0,1\}^{\ell(n)}}[\bm{D}(G_n(\bm{x})) = 1] - \Prob_{\bm{D},\, \bm{y} \sim \{0,1\}^n}[\bm{D}(\bm{y}) = 1] \; \right | \;\leq\; 1/n.
			\]
		\end{itemize}
	\end{hypothesis} 
	 	 
	Note that in \CryptoETH~ it is necessary to fool algorithms that run in time $2^{\gamma \cdot \ell(n)}$ in the \emph{seed length} $\ell(n)$. 	We say that  $\mathsf{Crypto}\text{-}\mathsf{ETH}$ holds if \CryptoETH~is true for some $\gamma > 0$, and that $\mathsf{Crypto}\text{-}\mathsf{SETH}$ holds if \CryptoETH~is true for every $\gamma  \in (0,1)$.

	The next result formalizes \Cref{thm:coding_lower_bound} from Section \ref{sec:intro_lb}.  It shows that if $\mathsf{Crypto}\text{-}\mathsf{ETH}$ holds then the best parameter achieved by an \emph{efficient} coding theorem for $\rKt$ is $(1 + \Omega(1)) \cdot \log(1/\delta) + \mathsf{poly}(\log n)$. On the other hand, if the stronger $\mathsf{Crypto}\text{-}\mathsf{SETH}$ hypothesis holds, then no efficient coding theorem for $\rKt$ achieves parameter $(2 - o(1)) \cdot \log(1/\delta) + \mathsf{poly}(\log n)$. 

\begin{theorem}[Conditional  $\rKt$ lower bound of efficient compression for poly-time samplers]\label{t:lbd}
		Suppose that \CryptoETH~holds for some constant $\gamma \in (0,1)$, and let $\ell(n)$ be the corresponding seed length function. There is a polynomial-time  sampler $S = \{S_n\}_{n \geq 1}$  such that, for every $\varepsilon > 0$ and $C \geq 1$, the following holds. For every probabilistic polynomial time algorithm $F$, there is a sequence $\{y_n\}_{n \geq 1}$ of strings $y_n \in \{0,1\}^n$ such that:
		\begin{itemize}
			\item[\emph{(}i\emph{)}] Each string $y_n$ is sampled by $S_n$ with probability $\delta_n \geq 2^{-\ell(n)}$.
			\item[\emph{(}ii\emph{)}] On some input parameter $\delta'_n  \leq \delta_n$, $F$ fails to output a probabilistic representation  of $y_n$ of  certifying an $\rKt$ complexity 
			\[
			k_n(\delta'_n) = (1 + \gamma - \varepsilon) \cdot \log(1/\delta'_n) + C (\log n)^C.
			\]
			More precisely, there is a sequence $\{\delta'_n\}_{n \geq 1}$ with $(1/2)\delta_n \leq \delta'_n \leq \delta_n$  such that
			\[
			\Prob_{\bm{F}}\left[\,\bm{F}(n,y_n, \delta'_n,\mathsf{code}(S))~\text{outputs an}~\rKt~\text{encoding of}~y_n~\text{of complexity}\;\leq\, k_n(\delta'_n)\,\right]\;\to_n\; 0.
			\]
		\end{itemize} 
	\end{theorem}
	
	We remark that the \emph{lower bound} on the seed length $\ell(n)$ present in \CryptoETH~is a consequence of  the $\poly(\log n)$ additive term in the definition of $k_n$ in \Cref{t:lbd}. This makes the negative result more robust. On the other hand, the \emph{upper bound} on $\ell(n)$ is needed in our argument when considering an arbitrary $\gamma \in (0,1)$.

	\begin{proof}
		We implement the strategy described in Section \ref{sec:techniques}. 
		
		Consider the function family $\{ G_n\}_{n \ge 1}$ of PRG's witnessing that \CryptoETH\, holds, where $G_n: \zo^{\ell(n)} \mapping \zo^n$. We fix a sufficiently large length $n$ (so that the forthcoming argument works), and for convenience we refer to $G_n$ simply as $G$.
	
	Let $Y=\{y_1,y_2,\dots,y_m\}$ be the strings in the support of the PRG $G$ and let $p_1,p_2,\dots,p_m$ be their corresponding probabilities. Also, let $\delta_i$ be the unique number of the form $2^{q}/2^{\ell(n)}$, where $q=0,1,2,\dots, \ell(n)$, such that $(1/2) p_i < \delta_i \leq p_i$.

		We define our sampler $S_n$ as follows. $S_n$ on input $1^n$ flips a coin $\ell(n)$ times obtaining a random string $z \in \zo^{\ell(n)}$ and next outputs $G(z)$. Thus, $S_n$ runs in polynomial time, and for each $i \in [m]$, $S_n$ generates $y_i$ with probability $p_i$.

		We must prove that for every choice of $\varepsilon > 0$ and $C \geq 1$, and for every probabilistic polynomial time algorithm $F$, there is a sequence $\{y_n\}_{n \geq 1}$ of strings $y_n \in \{0,1\}^n$ and a sequence $\{\delta'_n\}_{n \geq 1}$ of probability bounds $\delta'_n$ with the properties described above. Note that Item (\emph{i}) is true by the definition of the sampler.

Suppose there is a probabilistic polynomial-time algorithm $F$ that violates the assumption of the theorem.  Then for every $i\in[m]$, there is a set $E_{y_i}$ of valid probabilistic representations of $y_i$ certifying $\rKt$ complexity at most $k(\delta_i)$ such that
\[
\Prob_{\bm{F}}[\bm{F}\left(y_i, \delta_i\right)\in E_{y_i}]\geq \zeta,
\]
for some constant $\zeta>0$. We view the elements of $E_{y_i}$ as the \emph{good} fingerprints of $y_i$.  The sets $E_{y_i}$ are pairwise disjoint, because no element can be a good fingerprint of two strings. For each $i\in[m]$, let $x_{i,j}$, where $j\in[|E_{y_i}|]$, be the $j$-th element in $E_{y_i}$. Let
\[
C \eqdef \left\{x_{i,j} \,\middle|\, i\in[m] \text{ and } j \in [|E_{y_i}|]\right\},
\]
i.e., $C$ is the event that a fingerprint is good for some $y_i$. We define
\[
p_{i,j} \eqdef \Prob_{\bm{F},\, \bm{z} \sim \{0,1\}^{\ell(n)}}\left[G(\bm{z})=y_i \text{ and } \bm{F}\left(y_i, \delta_i\right) = x_{i,j}\right].
\]
Note that we have $p_i\geq p_{i,j}$, for every $i$ and $j$. Also, $C$ has probability at least $\zeta$ because
\[
\sum_{x_{i,j} \in C} p_{i,j} = \sum_{i \in [m]} \Prob_z [G(z) = y_i] \cdot \sum_{j \in [|E_i|]} \Prob_{F,z} [ F(y_i, \delta_i) = x_{i,j} \mid G(z = y_i)] \geq  \sum_{i \in [m]} \Prob_z [G(z) = y_i] \cdot \zeta = \zeta.
\]

We say that a fingerprint $x_{i,j} \in C$ is \emph{not short} if  \[
|x_{i,j}| \geq \log (1/p_i) - 2 \lceil \log (|x_{i,j}| + 1) \rceil - 2 - \log(2/\zeta).
\]
Let $\mathcal{E} \eqdef \left\{x_{i,j} \mid i \in [\ell], j \in [|E_i|], \, \textrm{$x_{i,j}$ is not short}\right\}$ (i.e.,  $\mathcal{E}$ is the event that a fingerprint is good and not short).

\begin{claim}\label{claim:shortlength}
    $\Prob_{z,\rc}(\mathcal{E}) \geq \zeta/2$.
\end{claim}

\begin{proof}[Proof of \Cref{claim:shortlength}]
Let
\[
C_{\textrm{pf}} \eqdef \left\{x'_{i,j} \,\middle|\, i\in[m] \text{ and } j \in [|E_{y_i}|]\right\}
\] 
be a prefix-free encoding for the strings in  $C$, where $x'_{i,j}$ is obtained from $x_{i,j}$ using the standard trick that inserts in front of $x_{i,j}$ its length written in binary with every bit doubled followed by $01$ to delimit this addition from $x_{i,j}$. Note that $|x'_{i,j}| = |x_{i,j}| + 2 \lceil  \log  (|x_{i,j}|+1 ) \rceil + 2$. 

Consider the set
\[
C_{\textrm{pf}}^{\prime} \eqdef \left\{ x'_{i,j} \in C_{\textrm{pf}} \,\middle|\, \left|x'_{i,j}\right| \leq  \log(1/p_i)-\log(2/\zeta)\right\},
\]
which is the prefix-free encoding of the complement (with respect to $C$) of the event $\mathcal{E}$.

Since $C_{\textrm{pf}}^{\prime}$ is a prefix-free set, we can use Kraft's inequality and we obtain 
\begin{align*}
	1 &\geq \sum_{x'_{i,j} \in C_{\textrm{pf}}^{\prime}} 2^{-\left|x'_{i,j}\right|}\\
	& \geq \sum_{x'_{i,j} \in C_{\textrm{pf}}^{\prime}} 2^{ -\log(1/p_i)+\log(2/\zeta)} \\
	&= \sum_{x'_{i,j} \in C_{\textrm{pf}}^{\prime}} p_i \cdot (2/\zeta)\\
	&\geq \sum_{x'_{i,j} \in C_{\textrm{pf}}^{\prime}} p_{i,j} \cdot (2/\zeta), 
\end{align*}
which implies
\[
\mu(C_{\textrm{pf}}) = \sum_{x'_{i,j} \in C_{\textrm{pf}}^{\prime}} p_{i,j} \leq \zeta/2.
\]
The event $\mathcal{E}$ has $\mu$-probability $\mu(C) - \mu(C_{\textrm{pf}}') \geq \zeta - \zeta/2 = \zeta/2$.
This ends the proof of \Cref{claim:shortlength}.
\end{proof}

\Cref{claim:shortlength}
means that with probability at least $\zeta/2$, over a random seed $z$ and the internal randomness of $F$, $G(z)$ outputs some $y_i$ and $F(y_i,\delta_i)$ gives a valid probabilistic representation $x$ of $y_i$ certifying $\rKt$ complexity $|x| + \log t_x$ at most $k(\delta_i)$ and its length $|x|$ is at least
\[
s \eqdef	\log(1/p_i) - \log(2/\zeta) - (2 \lceil \log (|x|+1) \rceil + 2).
\]

We can implement a distinguisher $D$ that, given $y\in\bool^n$, does the following.
\begin{enumerate}
	\item For every 
	\[
	\delta'_q :=2^q/2^{\ell(n)},
	\]
	where $q = 0, 1,2,\dots,\ell(n)$, $D$ runs $F\left(y,\delta'_q\right)$ (using the same randomness of $F$ for all the $\delta'_q$), and obtain a collection of encodings $S:=\{x_q\}_q$.
	\item $D$ outputs $1$ if both of the following conditions hold for \emph{at least one} $x_q\in S$:
	\begin{itemize}
		\item $|x_q|\leq k(\delta'_q)$.
		\item $x_q$ can be decoded (probabilistically) in $2^{(\gamma-\varepsilon/2)\cdot \ell(n)}$ steps and the decoded string is equal to $y$.
	\end{itemize}
	Again, for decoding, we can use the same randomness for all the $x_q$.
\end{enumerate}
		It is easy to see that the running time of $D$ is $2^{\left(\gamma-\Omega(1)\right)\cdot \ell(n)}$. Next, we argue that $D$ is a distinguisher for $G$.
		\begin{claim}\label{claim:distinguish}
			We have
			\[
			\;\Prob_{\bm{D},\, \bm{z} \sim \{0,1\}^{\ell(n)}}[\bm{D}(G(\bm{z})) = 1] \geq \zeta/3 = \Omega(1),
			\]
			and
			\[
			\;\Prob_{\bm{D},\, \bm{y} \sim 		\{0,1\}^{n}}[\bm{D}(y) = 1] = o(1).
			\]
		\end{claim}
		\begin{proof}[Proof of \Cref{claim:distinguish}]
		For the first item, note that from the discussion above, we have that with probability at least $\zeta/2$ (over a random $z$ and the internal randomness of $F$), $G(z)$ outputs some $y_i$ and for $\delta'_q=\delta_i$, $F(y_i,\delta'_q)$ will output some probabilistic representation $x_q$ of $y_i$ certifying $\rKt$ complexity 
		$|x_q| + \log t_{x_q}$ at most $k(\delta'_q)$ and  length $|x_q|$ at least
		\begin{align*}
			s &:= \log(1/p_i) - \log(2/\zeta) - (2 \lceil \log (|x_q|+1) \rceil + 2) \\
			&\geq \log(1/2 \delta_i)- \log(2/\zeta) - (2 \lceil \log (|x_q|+1) \rceil + 2) \\
			&\geq \log(1/\delta'_q)- 3\log(n).
		\end{align*}
		 Whenever we have such an encoding, we can decode probabilistically in time
		\[
			t_{x_q} \leq 2^{k(\delta'_q)-s} = 2^{(1+\gamma - \varepsilon) \log(1/\delta'_q) + C (\log n)^C - s} \leq 2^{(\gamma-\varepsilon/2)\cdot \ell(n)},
		\]
		(we have used $\ell(n) = (\log n)^{\omega(1)}$) and with error probability at most $1/3$.
		By the definition of $D$, we conclude that $D$ outputs $1$ with probability at least $(\zeta/2) \cdot (2/3) = \zeta/3$.

		We now show the second item. Fix any $\delta'_q\geq 1/2^{\ell(n)}$, where $q=0,1,\dots,\ell(n)$. We will show that
		\begin{equation}\label{eq:uniform_imcompress}
			\Prob_{\bm{F},\, \bm{\mathsf{Dec}},\, \bm{y} \sim 		\{0,1\}^{n}}\left[\left|\bm{F}(\bm{y},\delta'_q)\right|\leq k(\delta'_q) \text{ and } \bm{\mathsf{Dec}}\!\left(\bm{F}(\bm{y},\delta'_q)\right)=\bm{y}\right] = o\!\left(\frac{1}{\ell(n)}\right).
		\end{equation}
		Then the second item follows from a union bound. For the sake of contradiction, suppose \Cref{eq:uniform_imcompress} is false. Then by averaging, there exist circuits $F'$ and $\mathsf{Dec}'$, which are obtained by fixing the randomness of $\bm{F}$ and $\bm{\mathsf{Dec}}$ and by hard-wiring $\delta'_q$, such that
		\[
			\Prob_{\bm{y} \sim 		\{0,1\}^{n}}\left[\left|F'(\bm{y})\right|\leq k(\delta'_q) \text{ and } \mathsf{Dec}'\left(F'(\bm{y})\right)=\bm{y}\right] = \Omega\!\left(\frac{1}{\ell(n)}\right).
		\]
		However, this is not possible, because by a counting argument, the probability on the left side is at most
		\[
			\frac{2^{k(\delta'_q)}}{2^n} 
			= \frac{2^{(1+\gamma-\varepsilon) \log(1/\delta'_q) + C (\log n)^C}}{2^n}\leq
			\frac{2^{(1+\gamma-\varepsilon) \ell(n) + C (\log n)^C}}{2^n} \leq
			\frac{2^{(2-\varepsilon) \ell(n) + C (\log n)^C}}{2^n} \leq 
			2^{-\Omega(n)},
		\]
		where we used that 
		$\gamma < 1$, 
		$\ell(n) \leq n/2$, and $\delta'_q \geq 2^{-\ell(n)}$. This completes the proof of \Cref{claim:distinguish}.
		\end{proof}
		The theorem now follows from \Cref{claim:distinguish}.
	\end{proof}

	\subsection{Fine-Grained Complexity of Coding Algorithms for Poly-Time Samplers}\label{sec:fine-grained}

	We prove the results stated informally in~\Cref{t:twosidedinformal}. The $\rKt$-complexity of a string adds together the length of a compressed codeword and the logarithm of the time it takes to decompress the codeword. In $2$-sided-$\rKt$ complexity we also consider the time to compress the string.

		\begin{definition}[$2$-sided-$\rKt$]\label{d:boundcompress}
	A sampler $A$ admits coding with $2$-sided-$\rKt_\varepsilon$ complexity bounded by $\Gamma$ if there exists a pair of probabilistic Turing machines $(\comp, \dcomp)$ such that for all $n$, all $y \in \zo^n$, and all $\delta > 0$ satisfying $\Prob[A(1^n) = y] \geq \delta$, it holds with probability $1-\varepsilon$ (over the randomness of $\comp$ and the randomness of $\dcomp$) that
	\begin{itemize}
	    \item $\comp(y, \delta)$ outputs a string $x$ in $t_C$ steps,
	    \item $\dcomp(x)$ outputs $y$ in $t_D$ steps, and
	    \item $|x| + \log(t_C + t_D) \le \Gamma$.
	\end{itemize}
	Here, $\Gamma$ is a function of $n$ and $\delta$. In case $\varepsilon < 1/3$, we drop the subscript in the notation $\rKt_\varepsilon$.
	\end{definition}

	The following result follows  from~\Cref{t:compsamplable} in the same way as~\Cref{t:effrKt}. 
	\begin{corollary}[Formal statement of~\Cref{t:twosidedinformal}, (a)] \label{t:fullrKt-positive}
	Let $S$ be a polynomial-time sampler. Then, for every $\varepsilon > 0$, $S$ admits coding with $2$-sided-$\rKt_\varepsilon$ complexity bounded by   $2 \log (1/\delta) + O(\alpha(|x|, 1/\delta, \varepsilon))$, where the constant hidden in $O(\cdot)$ depends on $A$  and $\alpha(|x|, 1/\delta, \varepsilon)$ is the function from~\Cref{e:alpha}.
	\end{corollary}
		\begin{theorem}[Formal statement of~\Cref{t:twosidedinformal}, (b)] \label{t:lbd-full-rKt}
	Assume \CryptoETH~holds for some $\gamma \in (0,1)$. Then there is a polynomial-time sampler $S$ that, for any $\varepsilon > 0$ and any $C > 0$,  does not admit coding with $2$-sided-$\rKt_{1/7}$ complexity bounded by $k_n(\delta) := (1+\gamma- \varepsilon) \log (1/\delta) +  C(\log n)^C$.
	\end{theorem}
	\begin{proof}  The proof is similar to the proof of~\Cref{t:lbd}, but there are differences caused by the fact that in 2-sided-$\rKt$, $t_C$ and $t_D$ (i.e., the runtimes of compression and decompression) are random variables, whereas in $\rKt$, $t_p$ is a fixed value (being the maximum decompression time over all the probabilistic branches).
	
	The PRG $G$, the set $Y = \{y_1, \ldots, y_m\}$, and the sampler $S$ are exactly like in the proof of~\Cref{t:lbd}.

Suppose there is a pair $(\comp, \dcomp)$ of probabilistic Turing machines  that violates the conclusion of the theorem for the sampler $S$, i.e., the pair certifies that the sampler $S$ has coding with  $2$-sided-$\rKt_{1/7}$ complexity bounded by $(1+\gamma- \varepsilon) \log (1/\delta) + C (\log n )^C$ for some $\varepsilon  > 0$ and $C > 0$. We  show that this assumption implies the existence of a distinguisher $D$ that breaks the security of $G$ stipulated by
\CryptoETH, and this contradiction proves the theorem.

Since for every $i\in[m]$,
$\Prob_{\rc, \rd} [\dcomp (\comp(y_i, \delta_i)) = y_i] \geq (1-1/7)$,
where the probability is over the random coins $\rc$ of $\comp$ and $\rd$ of $\dcomp$, a simple argument that considers separately the random coins of the two algorithms implies that
that for every $i\in[m]$ there is a set $E_{y_i}$ of strings (which we view as the \emph{good} fingerprints of $y_i$) such that
\begin{itemize}
\item[(a)] $\Prob_{\rc}[\comp\left(y_i, \delta_i\right)\in E_{y_i}]\geq 5/7$, and
\item[(b)] for every $x \in E_{y_i}$, $\Prob_{\rd}[\dcomp(x) = y_i] > 1/2$.
\end{itemize}
By (b), the sets $E_{y_i}, i \in [m]$ are pairwise disjoint. For each $i\in[m]$ and $j\in[|E_{y_i}|]$, let $x_{i,j}$  be the $j$-th element in $E_{y_i}$. Let
$C \eqdef \left\{x_{i,j} \,\middle|\, i\in[\ell] \text{ and } j \in [|E_{y_i}|]\right\}$ (i.e., $C$ is the event that a fingerprint is good for some $y_i$).

Let $\zeta = 5/7$. 
Taking into account (a), it follows, exactly like in~\Cref{t:lbd}, that $\Prob_{z, \rc}(C)$ is at least $\zeta$. As before, we say  that a fingerprint $x_{i,j} \in C$ is \emph{not short}  if 
$|x_{i,j}| \geq \log (1/p_i) - 2 \lceil \log (|x_{i,j}| + 1) \rceil - 2 - \log(2/\zeta)$.
Let $\mathcal{E} \eqdef \left\{x_{i,j} \mid i \in [m], j \in [|E_i|], \, \textrm{$x_{i,j}$ is not short}\right\}$ (i.e.,  $\mathcal{E}$ is the event that a fingerprint is good and not short).

\begin{claim}\label{claim:shortlength-2}
    $\Prob_{z,\rc}[\mathcal{E}] \geq \zeta/2$.
\end{claim}
\begin{proof}[Proof of~\Cref{claim:shortlength-2}] Identical to the proof of~\Cref{claim:shortlength}.
\end{proof}

\Cref{claim:shortlength-2}
means that, conditioned on the event $\mathcal{E}$ (so with probability of $(z, \rc)$ at least $\zeta/2$), $G(z)$ outputs some $y_i$ and $\comp(y_i,\delta_i)$ outputs a good fingerprint $x$ that is not short.

We implement a distinguisher $D$ that, on input $y\in\bool^n$, does the following.
\medskip

For every $q=0,1, \ldots, \ell(n)$:
\begin{enumerate}
	\item Let $\delta'_q :=2^q/2^{\ell(n)}$.
	 \item D runs $\comp\left(y,\delta'_q\right)$ (using the same randomness $\rc$ of $\comp$ for all the $\delta'_q$), which outputs  a fingerprint $x_q$. 
	 \item $D$ runs $\dcomp$   on input $x_q$, which outputs $y'$.  As above, $D$ uses the same randomness $\rd$ of $\dcomp$ for all the $x_q$.
	 
	\item \label{a:iter} $D$ outputs $1$ (and exits the for loop) if 
	\begin{itemize}
		\item  $y' = y$ (i.e., $\dcomp(\comp(y, \delta_q')) = y$), and 
		\item $t_C + t_D$ is at most $2^{(\gamma-\varepsilon/2)\cdot \ell(n)}$, where $t_C$ is the number of steps executed by $\comp$ on input $(y, \delta'_q)$ and $t_D$ is the number of steps executed by $\dcomp$ on input $x_q$.
	\end{itemize}
	
\end{enumerate}

If the for loop ends without the conditions in~\Cref{a:iter} being satisfied at any iteration, $D$ outputs $0$.
\medskip

Since randomness is re-used at every iteration, overall, $D$ is using randomness $(\rc, \rd)$.

	We now argue that $D$ is a distinguisher for $G$ with runtime bounded by  $2^{\left(\gamma-\Omega(1)\right)\cdot \ell(n)}$, which yields the desired contradiction and finishes the proof.
	
	First, since the execution of $D$ consists of $\ell(n)+1$ iterations  and it can be arranged that each iteration takes at most $ 2^{(\gamma-\varepsilon/2)\ell(n)}$ steps (by halting the iteration when $t_C + t_D$ gets larger than this value), the runtime of $D$ is, as claimed,  bounded by  $2^{\left(\gamma-\Omega(1)\right)\cdot \ell(n)}$.  Next we show that $D$ distinguishes the distributions $G(U_{\ell(n)})$ and $U_n$.
		\begin{claim}\label{claim:distinguish-2}
			We have
			\[
			\;\Prob_{\rc,\rd, z \sim \{0,1\}^{\ell(n)}}[D(G(z)) = 1] \geq \Omega(1),
			\]
			and
			\[
			\;\Prob_{\rc,\rd, y \sim 		\{0,1\}^{n}}[D(y) = 1] < o(1).
			\]
		\end{claim}
		\begin{proof}[Proof of \Cref{claim:distinguish-2}]
		We start with the first inequality. By the discussion above, conditioned on the event $\mathcal{E}$ (which by~\Cref{claim:shortlength-2} has probability of $(z, \rc)$ at least $\zeta/2$), $G(z)$ outputs some $y_i$ and for $\delta'_q=\delta_i$, $\comp(y_i,\delta'_q)$ outputs a good fingerprint  $x_q$ that is not short. Since $x_q$ is a good fingerprint,  conditioned on an event $\mathcal{E}' \subseteq \mathcal{E}$, $\dcomp$ on input $x_q$ reconstructs $y_i$. By (b), $\mathcal{E'}$ has probability of $(z, \rc, \rd)$ at least $1/2 \cdot \Prob[\mathcal{E}] \geq \zeta/4$.
		 Recall that $(\comp, \dcomp)$ are assumed to certify that the sampler $A$ is compressible with  $2$-sided-$\rKt_{1/7}$ complexity bounded by $(1+\gamma- \varepsilon) \log (1/\delta)$. This means that conditioned by an event $\mathcal{V}$ that has  probability  of $(z, \rc, \rd)$ at least $1-1/7$, it holds that
\[
|x_q| + \log(t_C + t_D) \leq k_n(\delta'_q).
\]
		Now, conditioned by $\mathcal{E'}$, $x_q$ is not short, and so (like in~\Cref{t:lbd}) 
		\[
		|x_q| \geq s := \log(1/p_i) - \log(2/\zeta) - (2 \lceil \log (|x_q|+1) \rceil + 2) \geq \log(1/\delta'_q)- 3\log(n).
		\]
		
		By combining the above two inequalities, it follows that conditioned by $\mathcal{E'} \cap \mathcal{V}$, which has probability of $(z, \rc,\rd)$ at least $\zeta/4- 1/7$, it holds that
		\[
			t_C + t_D \leq 2^{k_n(\delta'_q)-s} \leq 2^{(\gamma-\varepsilon/2)\cdot \ell(n)}.
		\]
		By the definition of $D$, we conclude that $D$ outputs $1$ with probability greater than $\zeta/4 - 1/7 = \Omega(1)$ (recall that $\zeta = 5/7$).
		
		The second inequality is shown in the same way as the second inequality of~\Cref{claim:distinguish}. \end{proof}
		Thus,  $D$ is a distinguisher that contradicts that the PRG $G$ has the security stipulated by \CryptoETH, and this finishes the proof of \Cref{t:lbd-full-rKt}.
	\end{proof}

	\section{A Coding Theorem for \texorpdfstring{$\mathsf{pK}^{\mathsf{t}}$}{pKt} Complexity and Its Consequences}\label{sec:pkt}

\subsection{Optimal Coding Theorem for \texorpdfstring{$\mathsf{pK}^{\mathsf{t}}$}{pKt}}

    In this section, we prove our optimal coding theorem for $\pK^t$.
	\begin{theorem}[Reminder of \Cref{thm:pkt_coding}]\label{thm:source_coding_pkt}
		Suppose there is a randomized algorithm $A$ for sampling strings such that $A(1^n)$ runs in time $T(n)$ and outputs a string $x \in \{0,1\}^n$ with probability at least $\delta > 0$. Then
		\[
		\pK^t(x) \,=\,  \log(1/\delta) +  O\!\left(\log T(n)\right),
		\]
		where $t(n) =\poly\!\left(T(n)\right)$ and the constant behind the $O(\cdot)$ depends on $|A|$ and is independent of the remaining parameters.
	\end{theorem}

	For a function $H\colon\bool^{\ell}\to \bool^T$, we will sometimes identify it with a string $H\in\bool^{2^\ell\cdot T}$.
	
	\begin{lemma}\label{lemma:HSG}
	    For any $T\in\mathbb{N}$ and $\delta\in[0,1]$, there exists a family of functions
		\[
		\caH =\left\{H_w\colon\bool^{\ell}\to\bool^T\right\}_{w\in\bool^{k}}
		\]
		where $k=\poly(T)$ and $\ell= \log(1/\delta)+O(1)$ such that the following holds. Let $M\colon\bool^T\to \bool^{*}$ be a function computable in time $T$ and let $x\in\mathrm{Range}(M)$ be such that
		\[
		\Prob_{z\sim\bool^T}[M(z)=x] \geq \delta.
		\]
		It holds that
		\[
		\Prob_{w\sim\bool^{k}}\left[\text{$\exists\, v\in\bool^{\ell}$ such that $M(H_w(v))=x$}\right]\geq 2/3.
		\]
		Moreover, given $w\in\bool^{k}$ and $v\in\bool^\ell$, $H_w(v)$ can be computed in time $\poly(T)$.
	\end{lemma}
	\begin{proof}
	    Consider arbitrary $M$ and $x$. Let us call a function $H\colon\bool^{\ell}\to\bool^{T}$ \emph{good} (with respect to $M$ and $x$) if there exists some $v\in\bool^\ell$ such that $M(H(v))=x$. First note that a random $H\in\bool^{2^\ell\cdot T}$ is good with high probability.
		In particular, the probability that a random $H$ is not good is at most $(1-\delta)^{2^{\ell}}$, which is at most $o(1)$ for our choice of $\ell$.
		
		Next, we show that checking whether a given $H$ is good can be implemented as a constant-depth circuit. More specifically, note that given $M$ and $x$, and using oracle access to $H$, checking whether there exists some $v\in\bool^{\ell}$ such that $M(H(v))=x$ can be done in $\NP$. By the standard connection between the computation of an oracle-taking machine in $\PH$ and constant-depth circuits (see e.g., \cite{DBLP:journals/sigact/RossmanST15}), we get that there is an ${\sf AC}^0$ circuit of size at most $2^{\poly(T)}$ that takes $H$ as input and checks whether it is good.
		
		Now we will try to generate a good $H$ using a pseudorandom generator for ${\sf AC}^0$ circuits. It is known that there is a pseudorandom generator $G\colon\bool^r\to\bool^{N}$ that $(1/10)$-fools ${\sf AC}^0$ circuits on $N$ bits of size at most $s$, where the seed length $r$ is at most $\mathrm{polylog}(Ns)$. Moreover, given $z\in\bool^{r}$ and $i\in[N]$, the $i$-th bit of $G(z)$ can be computed in time $\poly(r)$ (see e.g., \cite{DBLP:journals/combinatorica/Nisan91, DBLP:conf/coco/TrevisanX13,DBLP:conf/coco/Tal17,DBLP:conf/approx/ServedioT19}). Let $N\vcentcolon=2^{\ell}\cdot T$ and let $s\vcentcolon= 2^{\poly(T)}$. We get a generator that takes $w\in\bool^{\poly(T)}$ and outputs a function $H_w\in \bool^{2^{\ell}\cdot T}$, such that with probability at least $1-o(1)-1/10>2/3$ over $w$, $H_w$ is good. Finally, note that given $w$ and $v$, we can compute $H_w(v)$ in time $\poly(T)$ because we can compute any single output bit of the generator in time $\poly(T)$.
	\end{proof}

	We are now ready to show \Cref{thm:source_coding_pkt}.
	\begin{proof}[Proof of \Cref{thm:source_coding_pkt}]
		Let us view $M\vcentcolon=A(1^n)$ as a function that takes $T\vcentcolon=T(n)$ random bits and outputs $x\in\bool^n$ with probability at least $\delta$. 
		
		By \Cref{lemma:HSG}, for at least $2/3$ of $w\in\bool^{\poly(T)}$, we get a function $H_w\colon\bool^{\ell}\to\bool^T$, where $\ell=\log(1/\delta)+O(\log T)$, with the property that there is some ``good'' $v\in\bool^{\ell}$ such that $M(H_w(v))=x$. Also, given $w$ and $v$, $H_w(v)$ can be computed in time $\poly(T)$. This means that for at least $2/3$ of $w\in\bool^{\poly(T)}$, there is some advice string $\alpha\in\bool^{\log(1/\delta)+O(\log T)}$, which encodes the number $T$, the code for $A(1^n)$, the code for computing $H_w$ using $w$, and some good $v$ (which could depend on $w$), such that using $\alpha$ together with $w$ we can recover $x$ in time $\poly(T)$. This implies that
		\[
		    \pK^{t}(x)\leq \log(1/\delta)+O(\log T),
		\]
		where $t\colon\mathbb{N}\to\mathbb{N}$ is such that $t(n)=\poly(T(n))$.
	\end{proof}

	\subsection{Application: An Unconditional Version of Antunes-Fortnow}\label{sec:AF}
	In this subsection, we prove an unconditional version of a result in \cite{DBLP:conf/coco/AntunesF09}, which is stated in \Cref{thm:AF-pKt-intro}. We start with some useful lemmas.

	\subsubsection{Useful Lemmas}
	The following lemma lower bounds the $\pK^t$ complexity of a string by its (time-unbounded) Kolmogorov complexity.
	\begin{lemma}\label{prop:K_pK}
		For every computable time bound $t\colon\mathbb{N}\to\mathbb{N}$, there is a constant $b>0$ (which depends only on $t$) such that for every $x\in\bool^*$,
		\[
		\K(x)\leq \pK^{t}(x) + b\log(|x|).
		\]
	\end{lemma}
	\begin{proof}

		Recall the following source coding theorem for (time-unbounded) prefix-free Kolmogorov complexity. There is a universal constant $c>0$ such that, if there exists a randomized algorithm $D$ that uses randomness chosen from a prefix-free set and that generates $x$ with probability $\delta$, then
		\[
		\K(x\mid D)\leq \log(1/\delta) + c.
		\]
		Fix a computable function  $t$ and a string $x$. Given the integers $n:=|x|$ and  $k\vcentcolon=\pK^{t}(x)$, consider the algorithm $D$ that randomly picks $w\in\bool^{t(n)}$ and $\caM\in\bool^{k}$, and then outputs whatever $\caM(w)$ outputs within $t(n)$ steps. Note that the random strings used by $D$ all have the same length and thus they form a prefix-free set as required by the above coding theorem. By the definition of $\pK^t$, $D$ will output $x$ with probability at least $\frac{2}{3\cdot 2^k}$.   Consequently, using the above source coding theorem and the fact that $D$ can be encoded using $O_t(\log(|x|))$ bits (because $k \le |x|+O(1))$, we obtain
		\[
		\K(x)\leq k + b\log(|x|),
		\]
		where $b>c$ is some constant that depends only on $t$.
	\end{proof}
	
	For technical reasons, we introduce the following measure.
	
	\begin{definition}
		For a time bound $t\colon\mathbb{N}\to\mathbb{N}$ and $x\in\bool^*$, define
		\[
		\pK_*^t(x)\eqdef \pK^t(x) + b\log(|x|),
		\]
		where $b>0$ is the constant from \Cref{prop:K_pK}.
	\end{definition}
	
	We define the following (semi-)distribution which will be a key notion used in the proofs later.
	\begin{definition}\label{def:mt}
		For a time bound $t\colon\mathbb{N}\to\mathbb{N}$, let $m^t$ be the distribution over $\bool^*$ defined as
		\[
		m^{t}(x)\eqdef2^{-\pK_*^{t}(x)}.
		\]
	\end{definition}
	
	\paragraph*{Equivalence between polynomial time on $m^{\poly}$-average and worst-case time using \texorpdfstring{$\pK^t$}{pKt}. }
	
	\begin{lemma}\label{l:average_on_m}
		For any algorithm $A$ and any computable time bound $t\colon\mathbb{N}\to\mathbb{N}$, the following are equivalent.
		\begin{enumerate}
			\item $A$ runs in polynomial time on average with respect to $m^{t}$.
			\item The running time of $A$ is bounded by $2^{O\left(\pK_*^{t}(x)-\mathsf{K}(x)+\log(|x|)\right)}$ for every input $x$.
		\end{enumerate}
	\end{lemma}
	\begin{proof}
		The proof follows closely that of {\cite[Theorem 4]{DBLP:conf/fct/AntunesFV03}}. Let $t_A(x)$ denote the running time of $A$ on input $x$.
		\paragraph*{($2 \Longrightarrow 1$). }
		Let $c>0$ be a constant such that $t_A(x)\leq 2^{c\cdot \left(\pK_*^{t}(x)-\mathsf{K}(x)+\log(|x|)\right)}$. We have
		\begin{align*}
			\sum_{x\in \bool^*} \frac{t_A(x)^{1/c}}{|x|}\cdot m^t(x)
			&\leq  \sum_{x} \frac{2^{\pK_*^{t}(x)-\mathsf{K}(x)+\log(|x|)}}{|x|}\cdot 2^{-\pK_*^{t}(x)} \\
			&\leq \sum_{x} 2^{-\mathsf{K}(x)}<1,
		\end{align*}
		where the last line follows from Kraft's inequality.
		\paragraph*{($1 \Longrightarrow 2$). }
		For $n,i,j\in\mathbb{N}$ with $i,j\leq n^2$, define
		\[
		S_{i,j,n} \eqdef \left\{x\in\bool^n \mid 2^i\leq t_A(x)\leq 2^{i+1} \text{ and } \pK_*^{t}(x)=j \right\}.
		\]
		Let $r$ be such that  $2^r\leq \left|S_{i,j,n}\right|\leq 2^{r+1}$. We claim that for every $x\in S_{i,j,n}$, 
		\begin{equation}\label{e:average_on_m_eq1_claim}
			\mathsf{K}(x)\leq r+O(\log n).
		\end{equation}
		To see this, note that given $i,j,n$, we can first enumerate all the elements in $S_{i,j,n}$, which can be done since $t$ is computable, and then using additional $r+1$ bits, we can specify $x$ in $S_{i,j,n}$.
		
		Now, fix $i,j\leq n^2$, and let $r$ be such that  $2^r\leq \left|S_{i,j,n}\right|\leq 2^{r+1}$. Then by assumption and by the definition of $S_{i,j,n}$, we have for some constants $\varepsilon,c>0$ (which may depend on $m^t$ and hence the time bound function $t$),
		\[
		c > \sum_{x\in S_{i,j,n}} \frac{t_A(x)^{\varepsilon}}{|x|}\cdot m^t(x) \geq 2^r \cdot \frac{2^{\varepsilon\cdot i}}{n}\cdot 2^{-j}= 2^{\varepsilon\cdot i +r-j-\log (n)},
		\]
		which yields
		\[
		\varepsilon\cdot i +r-j-\log (n) < c.
		\]
		By \Cref{e:average_on_m_eq1_claim}, this implies that for every $x\in S_{i,j,n} $,
		\[
		\varepsilon\cdot i \leq \pK_*^{t}(x) - \mathsf{K}(x) + O(\log n).
		\]
		Therefore, we have that for every $x\in S_{i,j,n}$,
		\[
		t_A(x) \leq 2^{i+1}\leq 2^{\varepsilon^{-1}\cdot \left(\pK_*^{t}(x) - \mathsf{K}(x) + O(\log n)\right)} = 2^{O\left(\pK_*^{t}(x)-\mathsf{K}(x)+\log(|x|)\right)},
		\]
		as desired.
	\end{proof}
	
	\paragraph*{A ${\sf P}$-samplable distribution that dominates $m^{\poly}$.}
	\begin{lemma}\label{l:P_samplable_dominates_m}
		For any polynomial $p$, there is a ${\sf P}$-samplable distribution $\caD$ that dominates $m^p$.
	\end{lemma}
	\begin{proof}
		First note that there is a universal constant $d>0$ such that for every $x\in\bool^n$, $\pK^{p}(x)\leq n+d$. For a polynomial $p$, we define a distribution $\caD$ over $\bool^*$ as follow:
		\begin{enumerate}
			\item Pick $n$ with probability $\frac{1}{n\cdot (n+1)}$.
			\item Pick uniformly at random $j\in[n+d]$.
			\item Pick uniformly at random $w\in \bool^{p(n)}$.
			\item pick uniformly at random $\caM\in \bool^j$.
			\item Run $\caM(w)$ for $p(n)$ steps and output whatever is on its output tape.
		\end{enumerate}
		By the definition of $\pK^t$, for every $x\in \bool^n$, $\caD$ outputs $x$ with probability at least
		\[
		\frac{1}{n\cdot(n+1)}\cdot 
		\frac{1}{n+d}\cdot\frac{2}{3}\cdot 2^{-\pK^{p}(x)}\geq \frac{m^p(x)}{|x|^{O(1)}},
		\]
		as desired.
	\end{proof}
	
	\paragraph*{$m^{\poly}$ dominates  ${\sf P}$-samplable distributions. }
	\begin{lemma}\label{l:main}
		For every ${\sf P}$-samplable distribution $\caD$, there is a polynomial $p$ such that $m^p$ dominates $\caD$.
	\end{lemma}
	\begin{proof}
		Let $M_{\caD}$ be a probabilistic algorithm and let $q$ be the polynomial such that $M_{\caD}$ outputs $x$ with probability $\caD(x)$ within $q(|x|)$ steps. Consider any $n\in\mathbb{N}$. Let $M$ be a sampler that, on input $1^n$, runs $M_{\caD}$ for $q(n)$ steps and outputs whatever is on its output tape. It is easy to see that $M$ runs in time $\poly(q(n))$. Also, for every $x\in\bool^n$, $M(1^n)$ outputs $x$ with probability at least $\caD(x)$. By the coding theorem for $\pK^t$ (\Cref{thm:pkt_coding}), we have, for some polynomial $p$ (which depends on the running time of $M$),
		\[
		\pK_*^{p}(x)\leq \log(1/\caD(x)) + O(\log n),
		\]
		which implies
		\[
		m^{p}(x)=2^{-\pK_*^{p}(x)}\geq \frac{\caD(x)}{|x|^{O(1)}}.
		\]
		This completes the proof.
	\end{proof}
	
	\subsubsection{Putting It All Together}
	
	\begin{theorem}[Reminder of \Cref{thm:AF-pKt-intro}]
		The following are equivalent for every language $L$.
		\begin{enumerate}
			\item For every ${\sf P}$-samplable distributions $\caD$, $L$ can be solved in polynomial time on average with respect to $\caD$.
			\item For every polynomial $p$, there exist a constant $b>0$ and an algorithm computing $L$ whose running time is bounded by $2^{O\left(\mathsf{\pK}^{p}(x)-\mathsf{K}(x)+b\cdot \log(|x|)\right)}$ for every input $x$.
		\end{enumerate}
	\end{theorem}	
	\begin{proof}~\\
		\noindent ($1 \Longrightarrow 2$). Let $p$ be any polynomial. By \Cref{l:P_samplable_dominates_m}, there exists a ${\sf P}$-samplable distribution $\caD$ that dominates $m^p$. By assumption, there is an algorithm $A$ that computes $L$ and runs in polynomial time on average with respect to $\caD$. Then by \Cref{fact:domination_implies_average}, $A$ also runs in average polynomial time with respect to $m^p$. Finally, by \Cref{l:average_on_m}, we have that the running time of $A$ is bounded by $2^{O\left(\pK_*^{p}(x)-\mathsf{K}(x)+\log(|x|)\right)}$ for every input $x$, as desired.\\
		
		\noindent  ($2 \Longrightarrow 1$). Let $\caD$ be any ${\sf P}$-samplable distribution. By \Cref{l:main}, there is a polynomial $p$ such that $m^p$ dominates $\caD$. By assumption, there is an algorithm $A$ that computes $L$ such that on input $x$, $A$ runs in time at most 
		\[
		2^{O\left(\pK^{p}(x)-\mathsf{K}(x)+ b\cdot \log(|x|)\right)} \leq 		2^{O\left(\pK_*^{p}(x)-\mathsf{K}(x)+ \log(|x|)\right)}.
		\]
		Then by \Cref{l:average_on_m}, $A$ runs in polynomial time on average with respect to $m^p$, which by \Cref{fact:domination_implies_average} implies that $A$ also runs in average polynomial time with respect to $\caD$, as desired.
	\end{proof}

		\section{Concluding Remarks and Open Problems}\label{sec:open_problems}

	Our results indicate that \Cref{thm:rkt_coding} might be optimal among \emph{efficient} coding theorems for $\rKt$, i.e., those that efficiently produce representations matching the existential  bounds.  In the case of $\pK^t$, the corresponding coding theorem (\Cref{thm:pkt_coding}) is optimal. 
	We have described a concrete application of \Cref{thm:pkt_coding} (\Cref{thm:AF-pKt-intro}). A second application appears in \cite{GKLO22}. In both cases, achieving an optimal dependence on the probability parameter $\delta$ is critical, and for this reason, the result from \cite{DBLP:conf/icalp/LuO21} is not sufficient. 
	
    Naturally, we would like to understand the possibility of establishing an \emph{unconditional} coding theorem for $\rKt$ with an optimal dependence on the probability parameter $\delta$. While the validity of $\mathsf{Crypto}\text{-}\mathsf{ETH}$ implies that no \emph{efficient} coding theorem with this property exist, we have an \emph{existential} coding theorem of this form under a derandomization assumption (\Cref{p:conditional_optimal_coding_rkt}). In the case of $\K^t$ complexity, it is known that an unconditional coding theorem with optimal dependence on $\delta$ implies that $\EXP \neq \BPP$ (see \cite[Theorem 5.3.4]{TroyLeeThesis}). However, the techniques behind this connection do not seem to lead to an interesting consequence in the case of $\rKt$ and $\rK^t$. Consequently, an optimal coding theorem for $\rKt$ might be within the reach of existing techniques.
    
    It would also be interesting to establish \Cref{thm:coding_lower_bound} under a weaker assumption, or to refute $\mathsf{Crypto}\text{-}\mathsf{SETH}$. A related question is the possibility of basing $\mathsf{Crypto}$-$\mathsf{ETH}$ on the existence of one-way functions of exponential hardness. Existing reductions are not strong enough to provide an equivalence between one-way functions and cryptographic pseudorandomness in the  exponential regime (see \cite{DBLP:conf/crypto/VadhanZ13, DBLP:journals/siamcomp/HaitnerRV13}).

    Finally, are there more applications of $\pK^t$ complexity and of \Cref{thm:pkt_coding}? Since this coding theorem is both optimal and unconditional, we expect more applications to follow.\\
    

	\noindent \textbf{Acknowledgements.} We are grateful to Bruno Bauwens for discussions and useful insights. M.~Zimand was supported in part by the National Science Foundation through grant CCF 1811729.~Z.~Lu and I.C.~Oliveira received support from the Royal Society University Research Fellowship URF$\setminus$R1$\setminus$191059 and from the EPSRC New Horizons Grant EP/V048201/1.

	\bibliographystyle{alpha}	
	\bibliography{references}	
	
	\appendix
	
		\section{Estimating the probability of sampling a given string}\label{s:estimate}
 
In~\Cref{thm:rkt_coding}, we are assuming that the compressor has both the code of the sampler $A$ and $\delta$ which estimates from below the probability $p_x$ with which the string $x$ is sampled. 

This seems redundant, because with the sampler $A$ and $x$ in her hands, the compressor can run $A$ and find a good estimation $\delta$ of $p_x$.  While this is true, there is a cost: assuming black-box access to the sampler, we need to run it $\Omega(1/p_x)$ times to get an estimation of $p_x$ within a constant multiplicative factor. This follows from the following fact, proved by Bruno Bauwens (private communication). (Note: We have chosen the multiplicative factor of $2$ for simplicity, it can be replaced with any positive constant).

\begin{proposition}\label{p:bruno}
Consider the following task: the input is a binary string $u$ of length $N$. By doing random probes in $u$, we want to find with probability $1-\epsilon$ a number $\tilde{p} \in (\frac{1}{2}p, \, 2 p)$, where $p$ is the fraction of $1$'s in $p$.

Then the number of probes has to be larger than $m_\varepsilon(p) := (1/p) \cdot ((1/2) \ln (1/\varepsilon)))$ (provided $p > 0$).
\end{proposition}
\begin{proof} 
Suppose there is an algorithm that does at most $m_\epsilon(p_u)$ probes for all $N$-bit strings $u$  and finds the estimation of $p_u$ (the fraction of $1$'s in $u$) with the required precision, and with probability $1-\epsilon$.  Let $u_1$ be a string $\in \zo^N$ that has $sN$ $1$'s, and $u_2$ be a string in $\zo^N$ that has $4s N$ $1$'s, where $s$ is some value in $(0,1/8)$.   If we read $m := m_\varepsilon(s)$ probes from $u_1$, the probability that all the probes turn out to be $0$'s is  at least $(1-s)^m$, which  is greater than $\epsilon$. The same happens if we read $m_\varepsilon(4s)$ probes from $u_2$. If we only probe $0$'s, the algorithm will perform in the same way for both $u_1$ and $u_2$. Since the intervals $(\frac{1}{2}s, 2s)$ and $(\frac{1}{2}4s, 2 (4s))$ are disjoint, the algorithm will make a mistake in one of the two situations, contradicting that the error probability for all strings is at most $\epsilon$. 
\end{proof}

We next show that the lower bound in~\Cref{p:bruno} is tight: there exists an algorithm that runs the sampler $(1/p) \cdot 8 \log(1/\varepsilon)$ times and estimates $p$ within the multiplicative factor of $2$.  We start with the following lemma.

	\begin{lemma}\label{l:learnpx}
	Let $x$ be an $n$-bit string and let $p_x$ be the probability that a sampler $A$ produces  $x$. We assume $p_x > 0$. Let $s = 4 \log (1/\varepsilon)$ for some parameter $\varepsilon > 0$. A \emph{success} is a run   of the sampler $A$ that produces $x$. Let $T$ be the number of times we run the sampler till there are $s$ successes. Let {$\mathcal E$} be the event
	\[
	(1/2) \cdot s \cdot (1/p_x) \leq T \leq 2 \cdot s \cdot (1/p_x).
	\]
	Then {$\mathcal E$} has probability $1-2 \varepsilon$.
	\end{lemma}
	\begin{proof}
	Let $T$ be the number of samplings till there are $s$ successes. The expected value of $T$ is 
	$\mu_T = s (1/p_x)$,
	because the expected number of samplings till each success is $1/p_x$ and $T$ is the sum of $s$ random variables with this expectation.  
	
	We use a known technique to obtain concentration bounds for the geometric distribution. We first estimate the probability that the second inequality in event $\mathcal{E}$ fails. This is  the probability that $T > 2 \mu_T$, which is equal to the probability of the event $\mathcal{A}$ = ``In $2\mu_T$ samplings the number of successes is $< s$.''  Let $Z$ be the number of successes in $2 \mu_T$ samplings. The expected value of $Z$ is 
	\[\mu_Z = 2 \mu_T \cdot p_x = 2 s (1/p_x) \cdot p_x = 2s.
	\]
	Then the second inequality in $\mathcal{E}$ fails with probability
	\[
	\Prob[\mathcal{A}] = \Prob[Z < s] = \Prob[Z < (1/2) \mu_Z] < e^{-(1/4) \cdot (2s/2)} = \varepsilon.
	\]
	(We have used the Chernoff bound $\Prob[Z < (1 - \delta) \mu_Z] \le e^{-(\delta^2 \mu_Z)/2}$, for $\delta = 1/2$.)
	
	We now estimate in the same way the probability that the first inequality in event $\mathcal{E}$ fails,  which is  the probability that $T < (1/2) \mu_T$, which is equal to the probability of the event $\mathcal{B}$ = ``In $(1/2) \mu_T$ samplings the number of successes is $> s$.''  Let $W$ be the number of successes in $(1/2) \mu_T$ samplings. The expected value of $W$ is 
	\[
	\mu_W = (1/2)\mu_T \cdot p_x = (1/2) s (1/p_x) \cdot p_x = s/2.
	\]
	Then the first inequality in $\mathcal{E}$ fails with probability
	\[
	\Prob[\mathcal{B}] = \Prob[W > s] = \Prob[W > 2 \mu_W] < (e/4)^{s/2} < \varepsilon.
	\]
	(We have used the Chernoff bound $\Prob[W > (1 + \delta) \mu_W] \le \bigg( \frac{e^\delta}{(1+\delta)^{(1+\delta)}}\bigg)^{\mu_W}$, for $\delta = 1$.)
	\end{proof}

	\noindent
	\begin{proposition}[Algorithm for estimating the probability with which a string is sampled]\label{p:estimatesamplingprob}
	Let $A$ be a sampler that produces strings of length $n$. There is an algorithm that on input an $n$-bit string $x$ that is sampled by $A$,  and $\varepsilon > 0$, has the following behaviour with probability at least $1-2\varepsilon$:
	\begin{itemize}
	    \item If $p_x \geq 2^{-n}$, then it calls the sampler at most $(1/p_x)\cdot 8 \log(1/\epsilon)$ times and returns a value $\tilde{p} \in \bigg[\frac{1}{2}p_x, 2p_x \bigg]$,
	    \item If $p_x < 2^{-n}$, then it calls the sampler at most $2^n \cdot 8 \log(1/\varepsilon)$ times and returns a value $\tilde{p} \leq 2^{-(n-1)}$.
	\end{itemize}
\end{proposition}
	\begin{proof}
	The algorithm runs as follows: 
	\medskip
	
\noindent	
	We run the sampler multiple rounds and we  halt when either
	\smallskip
	
	(a) the sampler has obtained $x$ (the ``success'' event) $s := 4 \log(1/\varepsilon)$ times, or
\smallskip

	(b)  in $2 s 2^n$ sampling rounds, the number of successes is less than $s$.
	\smallskip
	
	\noindent
	 In other words, we stop sampling immediately when we obtain the $s$-th success, or if  $2s 2^n$ samplings did not manage to do this.
	 
	 \noindent
	 Let $T$ denote the number of samplings. In case (a), the algorithm returns $\tilde{p} = s/T$, and, in case (b) it returns  $\tilde{p}=2^{-n}$.
	 \bigskip

	 The conclusion follows with an  analysis of the following cases, in which we condition on the event $\mathcal{E}$  from ~\Cref{l:learnpx} (which holds with probability $1-2 \varepsilon$).
	
	\begin{itemize}
	    \item  Suppose $p_x \geq 2^{-n}$. Then $T\leq 2s \cdot(1/p_x) \leq 2s \cdot 2^n$, and therefore case (a) holds. Then the algorithm returns $s/T \in \big(\frac{1}{2} p_x, 2p_x\big)$ (recall that we are conditioning on $\mathcal{E}$).
	    \item Now, suppose  $p_x < 2^{-n}$. Then the algorithm returns either $\tilde{p} = s/T \leq 2 p_x < 2 \cdot 2^{-n}$ (if case (a) holds), or $\tilde{p} = 2^{-n}$ (if case (b) holds). The bound on the number of calls follows because the algorithm never does more than $2s 2^n$ calls.
	\end{itemize}
	\end{proof}

\end{document}